# Thermal Transport in Nanostructures

Nuo Yang,[1] Xiangfan Xu,[1,3], Gang Zhang,[2a)] and Baowen Li[1, 3b)]

Email: [a)] zhanggang@pku.edu.cn [b)] phononics@tongji.edu.cn

[1]NUS-Tongji Center for Phononics and Thermal Energy Science, School of Physical Science and Engineering, Tongji University, 200092 Shanghai, China

[2]Key Laboratory for the Physics and Chemistry of Nanodevices and Department of Electronics, Peking University, Beijing 100871, P. R. China

[3]Department of Physics and Centre for Computational Science and Engineering, Graphene Research Center, National University of Singapore, Singapore 117542, Republic of Singapore

**Abstract**

This review summarizes recent studies of thermal transport in nanoscaled semiconductors. Different from bulk materials, new physics and novel thermal properties arise in low dimensional nanostructures, such as the abnormal heat conduction, the size dependence of thermal conductivity, phonon boundary/edge scatterings. It is also demonstrated that phonons transport super-diffusively in low dimensional structures, in other words, Fourier's law is not applicable. Based on manipulating phonons, we also discuss envisioned applications of nanostructures in a broad area, ranging from thermoelectrics, heat dissipation to phononic devices.

**I. Introduction**

In the past two decades, rapid developments in synthesis and processing of nanoscale materials have created a great demand for understanding of thermal transport in low dimensional nanomaterials. Nanostructures include: one-dimensional (1D) structures, like nanotubes (NTs)[1] and nanowires (NWs);[2] two-dimensional (2D) crystal lattice with only one-atom-thick planar sheets, like graphenes;[3] and thin films consisting of alternating layers of two different materials, superlattices.[4,5]

To improve the speed of processors, the size of Metal-Oxide-Semiconductor Field-Effect Transistors (MOSFETs) has been scaled down to a gate length of 20 nm, and much smaller in the future. High intensity MOSFETs may induce more hot spots, so that the heat removal in microelectronics is a burning issue that limits the performance and reliability.

Nanostructures with high thermal conductivity can be applied in heat dissipation. To speed up applications of nanostructure in thermal management, a comprehensive investigation combining theoretical modeling, computational simulation, and experimental studies is indispensable.

In addition to great interests in applications, there is also a demand for the understanding of heat conduction in low dimensional systems. The conduction of heat is one of the most important fundamental physical phenomena in nature. Traditionally, heat flux through a homogeneous material is believed to follow the Fourier's law which states that heat flux is proportional to temperature gradient along the path of heat flow, as $J = -\kappa \cdot \nabla T$, where κ is thermal conductivity. For bulk material, the thermal conductivity is believed to be an intensive property which should be independent on the size and geometry, just as other material properties such as density, specific heat and electrical conductivity. Fourier's law has been confirmed in last two hundred years in bulk materials (three-dimensional macroscopic systems).

However, in the last few decades, it was found that the Fourier's law was violated in low dimensional lattices.[6-10] It was found that thermal conductivity of Fermi-Pasta-Ulam (FPU)-like chains diverged with the system size L, as $\kappa \propto L^{\beta}$. The reported value of β varies from 0.33 to 0.44. There is still an outstanding debate on a universal exponent value of β.[9,11] Wang and Li found that β equals 1/3 at intermediate coupling, and equals 2/5 at low temperature and weak coupling in 1D lattice. They also suggest that β is 1/3 when there is coupling between longitudinal and transverse modes.[7] Thus nano materials are very promising platforms to verify the fundamental thermal transport theory in low-dimensional systems.

In recent years, some theories have been developed to describe the heat transport in nanomaterials.[12-16] Various simulation approaches, such as molecular dynamics (MD),[17-21] Monte Carlo (MC),[22] lattice dynamics (LD),[23] non-equilibrium Green's function (NEGF),[24] Boltzmann transport equation (BTE)[25] etc., have been used in investigating nanoscale thermal transport. Furthermore, as advances in micro- and nano-technology, more and more experimental measurements of thermal properties of nanomaterials have been reported. The combination of theoretical, simulating and experimental investigations will definitely expedite our understanding of heat conduction in nanoscale.

In this article, we give a detailed discussion on present understanding of thermal transport in nanostructures. In Sec II, we show new physics and some novel thermal properties in 1D and 2D nanostructures, including size effect, boundary/edge scatterings and phonon coherent resonance effect. Sec III introduces the experimental works on nanotubes, nanowires and graphenes. We also discuss the applications of nanostructures in thermal management, phononic devices and thermoelectrics in Sec IV. At the end, we give conclusions and outlook in Sec V. Due to the length limit, we only address some important fundamental issues here, specially works from China. For comprehensive reviews on thermal properties of nanomaterials, please also refer to Refs. [10,26-35].

**II. Fundamentals of Thermal Transport in Nanostructures**

**A. 1D nanostructures**

Carbon nanotubes[36] (CNT) are one of the promising nanoscale materials discovered in the 1990s. They have many exceptional physical and chemical properties. From about ten years after it was discovered, the thermal property of nanotubes came to the spotlight of research, and attracted worldwide research interests.[1,37-62] Kim *et al.* observed the conductivity of multi-wall carbon nanotubes(MWCNs) is more than 3000 W/m-K at room temperature.[38] The ultra-high thermal conductivity of carbon nanotubes is from the pure honeycomb structure.

Isotope doping is an efficient way to reduce the thermal conductivity. The isotope effects on thermal conductivity of carbon nanotubes have been investigated by MD simulations,[45] and it is found that the thermal conductivity of carbon nanotubes can be reduced more than 50% with isotope impurity. Later, this large reduction by isotopic doping effect has been confirmed experimentally.[63]

Zhang and Li calculated the temperature dependence of thermal conductivity of (5,5) single-wall carbon nanotubes (SWCNTs).[45] Fig.1 shows the length dependent thermal conductivity of SWCNT at 300K and 800K, respectively.[45] It is clear that thermal conductivity $k$ diverges with nanotube length $L$ as $\kappa \sim L^{\beta}$, and β decreases from 0.40 to 0.26 as temperature increases from 300K to 800K. At high temperature, the transverse vibrations are much larger than that at low temperature, thus interaction between the transverse modes and longitudinal modes becomes stronger, which leads to a smaller value of β. The similar length dependent thermal conductivity is also observed in (8,8) SWCNTs.[1] This length dependent thermal conductivity was experimentally verified by Chang *et al.* in multi-wall carbon nanotubes, with β varies from 0.6 to 0.9 at room temperature.[44] These results demonstrate anomalous heat conduction in nanotubes.

To understand the mechanism of the length dependent thermal conductivity, the heat diffusion in SWCNT was studied.[64] Quantitatively, the width of the energy pulse can be measured by its second moment $\sigma^2(t) = \sum_i [E_{i,t} - E_0](\vec{r}_{i,t} - \vec{r}_{i,0})^2 / \sum_i [E_{i,t} - E_0]$, where $E_{i,t}$ is the energy of atom i at time t, $\vec{r}_{i,t}$ is the position of atom i at time t, and $\vec{r}_{i,0}$ is the position of energy pulse at t=0. The averaged energy profile spreads as $\langle\sigma^2\rangle \propto t^\alpha$. In Fig. 1 (c) and (d) we show $\langle\sigma^2\rangle$ versus time in logarithmic-logarithmic scale, so that the slope of the curve gives the value of α. It is clearly seen that at 2 K, energy transport ballistically in SWCNT; whereas at 300 K, energy transport super-diffusively with α=1.2. This is slower than ballistic transport, α=2, but faster than normal diffusion, α=1. At low temperature, the vibrations of atoms are very small, the potential can be approximated by a harmonic one, thus phonons transport ballistically in SWCNT. However at room temperature, the anharmonic terms appear due to excitation of the transverse vibrational mode, the situation changes dramatically.

In 2003, Li and Wang proposed a quantitative connection between heat conduction and diffusion.[6] By using the fractional derivative, they derived a relationship between β and α as $\beta = 2 - 2/\alpha$, which establish a connection between the microscopic process and the macroscopic heat conduction. They stated that an anomalous diffusion indicates an anomalous heat conduction which corresponds to a divergent thermal conductivity. If we use $\alpha = 1.2$ for SWCNT, then $\beta = 2 - 2/\alpha = 1/3$, which is in good agreement with the direct MD simulation. The anomalous heat diffusion is responsible to the length dependent thermal conductivity.

In addition to CNTs, Silicon nanowires (SiNWs) have attracted a great attention in recent years because of their excellent physical and chemical properties, and their potential applications in many areas including biosensor,[65,66] electronic device,[67,68] and solar photovoltaics.[69,70] SiNWs are appealing choice due to their ideal interface compatibility with conventional Si-based devices.

Because of boundary scatterings from high surface to volume ratio and confinements on number of phonon modes, Volz and Chen predicted that thermal conductivity of SiNW is about two orders of magnitude smaller than that of bulk Si.[2], which is confirmed experimentally by Li *et al*. later.[71] In 2005, a Monte Carlo method is developed to simulate phonon transport in NWs, which is in good agreement with the experimental data.[72] The reduction in thermal conductivity was attributed to the phonon group velocity reduction and the phonon lifetime reduction due to strong phonon-phonon scattering. Moreover, because of the large surface to volume ratio,[73,74] the boundary scattering in quasi-1D structure is also significant.[75,76] It was demonstrated that the disordered surfaces and decreased lifetimes of propagating modes are responsible for the reduction of thermal conductivity in SiNWs.[77] The impacts of size, cross-section, interface, defect, and surface adsorption on thermal conductivity of nanowire have been studied.[17,22,78-89]

By using equilibrium and non-equilibrium MD simulation, Schelling *et al*. reported the finite-size effect on thermal conductivity of SiNWs.[80] They showed that the results obtained by the equilibrium and non-equilibrium methods were in reasonable agreement with experimental results of SiNWs. Later, Liang and Li derived an analytical formula of diameter dependent thermal conductivity of NWs, where the surface scattering and the size confinement effects of phonon transport are considered.[81] The thermal conductivity predicted by their analytical formula were in good agreement with experimental data for SiNWs, GaAs NWs and thin films.

In 2010, Yang, Zhang and Li demonstrated the length dependent thermal conductivity and the anomalous heat diffusion in SiNW by using non-equilibrium MD.[17] As shown in Fig.2 the thermal conductivity increases with NW length up to 1.1 μm, and the length dependence of thermal conductivity is different in two length regimes. At room temperature, when length is less than about 60 nm, the thermal conductivity increases with the length linearly ($\beta = 1$). For the longer wire, the diverged exponent reduces to 0.27. In addition, the diverged exponent also depends on temperature. At 1000 K, $\beta$ is only about 0.15 when L>60nm. This critical length, 60nm, is in good agreement with previous predicted value of mean free path in bulk Si.[90] When SiNW length is less than the phonon mean free path, the phonon-phonon interaction can be neglected, then the phonons transport ballistically like in 1D lattice of harmonic oscillators, and the thermal conductivity increases with the length linearly.

When the length of SiNW is much longer than mean free path, the phonon-phonon scattering plays a key role in the process of phonon transport. Different from 3D bulk material, phonon-phonon interaction alone is not sufficient to reach a diffusive process in 1D SiNW, that is, the phonons transport super-diffusively which results in a diverged thermal conductivity. The results demonstrate that the SiNW is a promising platform to verify phonon transport mechanisms in low-dimensional systems. The numerical results demonstrate heat conduction of SiNW does not obey Fourier's law even though the NW length reach micro-scale, and the super diffusion is responsible for the length dependent thermal conductivity.

Fourier's law has received great success in describing macroscopic heat transport. The belief is for a certain material with definite composition and structure, the thermal conductivity is an intensive property that should be independent on the geometry and size, just as other intrinsic properties such as density, specific heat and electric conductivity. This characteristic has been confirmed by two hundred years of experimental observations. However, a rigorous proof for Fourier's law in micro/nano-scopic Hamiltonian dynamics is still lacking. It is still an open and debated question whether Fourier's law is valid in low dimensional systems. From the length dependent thermal conductivity, it is concluded that thermal conduction is anomalous in 1D nano materials. This anomalous phenomenon has been referred by some people as "breakdown of Fourier's law" in literature.

For nanomaterials, the natural defects and isotopic doping in the process of fabrication affect the thermal transport properties a lot. Doping with isotopes atoms increases phonon scattering, thus results in reduction of thermal conductivity.[29] In 2008, Yang, Zhang and Li investigated the isotopic effect on thermal conductivity of SiNWs.[79] As shown in Fig.3, the thermal conductivity of SiNWs can be reduced exponentially by random isotopic doping and reach a minimum value, 50% doped, as about 27% of that in pure $^{28}$Si NW. The heavier isotope atoms can decrease conductivity much more than the lighter ones. More interesting, the thermal conductivity of isotopic-superlattice structured SiNWs depends obviously on the period of superlattice as shown in Fig.3. The superlattice structure can reduce the conductivity significantly because the mismatch in power spectra of isotropic atoms with different mass. At a critical period of 1.09 nm, the thermal conductivity is only 25% of the value of pure SiNWs. More interestingly, the thermal conductivity increases anomalously when the superlattice period length is smaller than this critical value. This anomalous increase in thermal conductivity was explained by the collective vibrations of different mass layers in superlattice structured SiNWs when superlattice period is smaller than this critical value. The similar periodic length dependent thermal conductivity was also observed in Kr/Ar superlattice NWs.[78]

In addition to the random and interface scatterings of phonon, the surface scattering is another way to reduce thermal conductivity. It has been demonstrated that thermal conductivity of NWs can be reduced further obviously by introducing more surface scattering: making SiNWs hollow to create inner surface, i.e. silicon nanotubes (SiNTs).[88] Fig. 4 (a) shows the room temperature thermal conductivity of SiNWs and SiNTs versus cross section area. Even with a very small hole, only a 1% reduction in cross section area induces the reduction of thermal conductivity of 35%. Moreover, with increasing size of the hole, a linear dependence of thermal conductivity on cross section area is observed. The reduction of thermal conductivity can be understood from the analysis of phonon participation ratio (p-ratio) (Fig. 4 (b)). There is a reduction of p-ratio in SiNTs for both low and high frequency phonons, compared with SiNWs. For those localized modes with p-ratio less than 0.2, it is clear that the intensity of localized modes is almost zero in the centre of NW, while with finite value at the boundary (Fig. 4 (c)). This demonstrates that the localization modes in SiNWs are distributed on the boundary (especially at the corner) of cross section plane. In addition, due to inner-surface introduced in SiNTs, energy localization also shows up around the hollow region (see Fig. 4 (d)).

Most of the above approaches to reduce thermal conductivity, such as introduction of rough surface and defect scatterings, are based mainly on incoherent mechanisms, which cause phonons to lose coherence. In the following, we discuss how to use the phonon coherent resonance to tune thermal conductivity.[75]

The coherence of phonons in Ge/Si core-shell NWs, SiNWs and SiNTs can be quantitatively described by the heat current autocorrelation function (HCACF). For both SiNWs and SiNTs, there is a very rapid decay of HCACF at the beginning, followed by a long-time tail with a much slower decay to zero. The long-tail of HCACF reveals that this nontrivial oscillation is not random but shows a periodic manner. As is shown in Fig. 5 (c), the resonant amplitude first increases to a peak value and then decreases as size of core increasing. Interestingly, the resonant amplitude at different temperature shows the same structure dependence. To understand the mechanism, Fig. 5 (d) shows the fast Fourier transform (FFT) of normalized HCACF for core-shell NWs, which are very similar to the spectrum of the coherent resonance effect of in a confined structure. Atoms near the core-shell interface are stretched due to the different sound velocity, which induces a strong coupling between the longitudinal and transverse modes. This coupling is weakened in pure SiNWs due to the same sound velocity of atoms. This coupling picture gives explanation to the coherent resonance effect in the transverse direction can indeed manifest itself in HCACF along the longitudinal direction in core-shell NWs. Moreover, when longitudinal phonons transport along core-shell NWs, their energy are dispersed to the transverse direction due to the coherent resonance and mode coupling.

Very recently, the direct connection between coherent resonance and thermal conductivity reduction was reported[91] in Ge/Si core-shell. Fig. 6 (a)-(b) compares the p-ratio for eigen-modes in pristine GeNWs, Ge/Si core-shell NWs with perfect interface, and Ge/Si core-shell NWs with rough interface. There is a reduction of p-ratio in low frequency regime for coated NW, which is caused by the resonance induced by coupling between the transverse and longitudinal modes. As the strongest resonance peak is related to the eigen-mode with lowest frequency in transverse direction, the localization is remarkable for phonons with long wavelength (>NW diameter). The p-ratio in core-shell NWs shows an overall reduction for very low frequency longitudinal acoustic (LA) phonons (Fig. 6 (c)). An obvious dip of the p-ratio is observed at around 0.5 THz, which is consistent with the resonance frequency found in FFT of HCACF along the longitudinal direction and thus provides the strong evidence that phonon resonance in core-shell NWs leads to localization for low frequency phonons. More importantly, core-shell induces localization for low frequency phonons, while interface roughness localizes the high frequency phonons (Fig.6 (b)). When the coating thickness is less than certain critical value, thermal conductivity of Ge/Si core-shell NWs is smaller than that of pristine GeNW (Fig. 6 (d)). Thus the core-shell structure offers the unique opportunity to further reduce thermal conductivity of low thermal conductivity material even by coating with high thermal conductivity material. This offers novel avenues for the design and thermal management in nanostructures. Experimentally, the reduction in thermal conductivity in core-shell NW has been verified.[92]

**B. 2D nanostructures**

In addition to the one-dimensional nanomaterials, graphenes[93,94] and narrow graphene nanoribbon (GNR) have attracted immense interests recently, mostly because of their exciting physical properties caused from the unusual one-atom-thick structure. Superior thermal conductivity (as high as 5000 W/m-K) has been observed in graphene,[3] which has raised the exciting prospect of using them for thermal management devices. There are rich physical phenomena about thermal property of GNRs. The effects of size,[95-98] defects,[99,100] doping [101,102], shape,[103,104] stress/strain,[105-108], substrates,[109] inter-layer interactions,[110-112] nanoscale junctions,[113] chirality,[114] topological structure,[115] edge effect,[116-119] foldings (gradfolds),[23,120] etc. on thermal conductivity of nanoribbons[121-128] have been widely studied.

Due to different boundary condition, intrinsic anisotropy thermal conductivity of GNR is reported.[114] As shown in Fig. 7 (a), the room temperature thermal conductance of zigzag GNR is about thirty percent larger than that of armchiar GNR. This anisotropy phenomenon will disappear when the width is larger than 100 nm. For both armchair and zigzag GNR, its thermal conductivity depends on the width.[118] As shown in Fig. 7 (b), the thermal conductivity of zigzag GNR increases firstly and then turns to decrease with the width increasing, while the armchair GNR's thermal conductivity monotonously increases with the width increasing, which comes from the competition between the edge localized phonon effect and phonons' Umklapp effect. An interesting resonant splitting of phonon transport in periodic T-shaped GNR (Fig. 7 (c)) was observed by using NEGF method.[129] These resonant peaks are originated from high quasi-bound states in which phonons are mainly localized in the constrictions. And temperature dependent thermal conductivity (Fig. 7 (d)) of single-layer and multi-layer GNRs are also investigated by using MD simulations.[111]

Zhai and Jin investigated the strain effect on ballistic thermal transport in GNR by combination of NEGF and the elastic theory.[106] At temperature below 50 K, the thermal conductance increases observably for GNR with stretching strain of 10% and 19%, due to a lot of dispersive phonon modes are converged to the low frequency region. It should be noted the structure become unstable when strain increases from 19% to 20%. However, at high temperature, the strain effect is totally different: The room temperature thermal conductivity of GNRs decreases up to 60% under tensile strain, due to the softening of the phonon modes.[107]

The edge modes play an important role in thermal conductivity of GNR. Jiang and Wang[116]

predicted that there exist phonon localized edge-modes in 2D square lattice and graphene when two conditions are satisfied simultaneously: (I) couplings between different directions in the interaction; (II) different boundary condition in three directions. As shown in Fig. 8 (a) and (b), by applying the open boundary condition in the x and periodic boundary condition in y directions, both conditions (I) and (II) are satisfied and there is localized edge-modes. In Fig. 8 (c) and (d), the condition (II) is broken by applying open boundary condition in both x and y directions. Now there are only extended modes in this system, no localized edge-mode.

As a combination of stress/strain, inter-layer and substrate effect, the folded GNR gives rise to more interesting physics, because of their fascinating electrostatic and electronic properties. The folded GNR is also a good candidate and a promising way for future phonon engineering in graphene derivatives.[23,120] In 2012, Yang et. al have investigated thermal transport in folded GNRs systematically using MD, LD, and NEGF method. As shown in Fig. 9 (d), the percentage of reduction is dependent on the number of folds. Moreover, the more the structure with folds is compressed, the more the thermal conductivity is reduced. The thermal conductivity of GNR with 6 folds can be substantially decreased up to 70% compared to the flat GNR. The transmission spectra show the decrease of thermal conductivity comes from strong scattering of low frequency modes at the folds (Fig.9 (b) and (c)), which is much different from other high frequency phonon scatterings by impurities, dislocation, and boundaries. In addition, compressing the interlamellar space provide more number of phonon states available for three-phonon scatterings. The results suggest that besides geometries, sizes and other modulation methods, the method of folding provides additional freedom of manipulation transport properties in graphene, which may further strengthen its position as a mainstream building block in future device fabrication.

**III. Thermal Measurements in Nanoscale**

**A. 1D nanostructures**

The achievements of modern nano-fabrication technology have enabled us to measure thermal properties of nanomaterials, although still with great challenges. Thermal conductivity of multi-walled carbon nanotubes (MWCNTs) bundles was firstly measured to have relatively low values, probably due to the dominant scatterings in the barriers between tubes.[130-132] In order to reduce the inter-tube scatterings and to study the intrinsic thermal conductivity of individual MWCNT, Kim *et al*.[38] adopted the thermal-bridge method by integrating complex electron beam lithography and nano-manipulation. A suspended microelectromechanical system (MEMS) device was used to measure the thermal conductivity of MWCNT. Unlike the traditional method using a bundle of MWCNTs in the form of rod or filament, an individual MWCNT was measured. As shown in Fig. 10 (upper insert), Rh and Rs are Pt loop on top of suspended SiNx membranes. A Joule heat is created by applying electrical current in $R_h$, which will dissipate through both the three supporting $SiN_x$ beams and the MWCNT, creating a temperature raise in $R_s$. Under steady state, thermal conductivity of interested sample can be calculated by measuring the temperature raises in $R_h$ and $R_s$. The observed thermal conductivity is more than 3000 W/m-K in MWCNT with diameter of 14 nm, which is two orders of magnitude higher than that in MWCNTs bundles.

Following the microfabricated structures mentioned above, significant progress has been made in investigating the thermal conductivity in various one dimensional structures, such as SiNWs,[133-135] CNTs and boron nitride nanotubes,[44,63,136-138] ZnO NWs,[139] $VO_2$ nanobeams,[140] etc.. Very recently, Chen *et al*. developed this technique with a dramatic improved sensitivity, resulting in the capability to verify the theoretically predicted phonon coherent resonance effect[75] in Ge-Si core-shell ultrathin NWs.[141,142]

The development of suspended microfabricated structures for thermal properties measurement has paved the way for many significant studies in understanding the fundamental thermal transport at nanoscale.[44,71,133] Recently, by taking advantage of their advances in nanofabrication, Li and Thong managed to fabricate the MEMS in the size of 8 inches, with more than 600 devices integrated in one single wafer (insert in Fig. 11).[139,140,143]

Based on this technology, a systematic study of the thermal conductivity of individual single-crystal ZnO NWs with diameters ranging from 50nm to 210nm was performed by Li and Thong's groups.[139] The thermal conductivity of ZnO NWs is found to be significantly reduced when comparing to that in bulk materials, due to the enhanced phonon-boundary scatterings. An empirical relationship for diameter-dependent thermal conductivity was demonstrated, which shows an approximately linear dependence of thermal conductivity with cross-section area (Fig. 11). This diameter dependence is expected to stimulate further theoretical investigations of the intrinsic effect of size on the thermal transport in NWs.

Understanding thermal transport in nanomaterials, more specifically the interaction between nanostructrues, are of intriguing interests for developing thermal management and thermal conversion applications.[54,138,144] Recently, thermal conductivity of double boron nanoribbons (coupled with van der Waals interaction) was found to be significantly higher than that of free-standing individual nanoribbons (Fig. 12 (a)).[138] This observation indicates that a significant portion of phonons transport the interface of the two boron nanoribbons without being scattered.

In additional, Yang *et al*. also found double nanoribbons prepared in IPA solution has a relative lower thermal conductivity than that of an individual nanoribbon (Fig. 12 (b) ).[138] The authors then wetted the same double nanoribbons in a reagent alcohol and DI water mixture and found that the thermal conductivity increased by ~15%-20%. This enhancement could be eliminated by wetting the nanoribbons again in IPA (Fig. 12 (b)). This provides a simple approach to tune thermal conductivity of bundle nanomaterials coupled with van de Waals interface.

Despite the significant progresses and successes in measuring thermal conductivity of nanomaterials using the suspended MEMS, there are still many challenges. The primary challenge lies in the difficulties in measuring the thermal contact resistance at the two ends of sample, which would have major contribution to the total measured thermal resistance. To this end, Yang *et al*. carried out a systematical measurement on length-dependent thermal resistance in MWCNT and found that the thermal contact resistance could contribute up to 50% of the total measured thermal resistance.[54,144] This indicates that the commonly used electron-beam induced metal deposition at the contacts may not reduce the thermal contact resistance to a neglected level.[144] More efforts need to be put to eliminate the thermal contact resistance, or measure it out directly.

**B. 2D nanostructures**

Although theories have uncovered the intriguing thermal conductivity in 2D systems, there is little progress on experiment till the discovery of graphene in 2004.[93,94,145] Graphene provides a perfect test field for studying the thermal transport in 2D systems. The pioneering experimental studies of thermal conductivity in suspended graphene were carried out based on Raman spectroscope measurements.[3,146-148] The thermal conductivity of exfoliated graphene was found to reach a high value of ~4800-5300 W/m-K at room temperature.

Several following experiments on graphene based on Raman measurements confirmed its ultra high room temperature thermal conductivity, which ranging from ~1800W/m-K to

~5000W/m-K.[149-152] Similar to the isotopic effect on thermal conductivity of CNT and SiNW, isotopic doping reduces thermal conductivity of graphene from 2805W/m-K to 2010 W/m-K with 1.1% $^{13}$C concentration.[153]

Besides the Raman based thermal measurements on suspended graphene, the traditional thermal-bridge method mentioned above, was also adopted to measure thermal conductivity.[143,154] Fig. 13 (a) and (b) show the SEM images of suspended and supported graphene. Fig. 13 (c) presents the thermal conductance per cross section area ($\sigma$/A) of graphene sheet with 500nm in length and 3$\mu$m in width.[154] Interestingly, the data at low temperatures can be fitted by $1.8 \times 10^5 \times T^{1.5}$ W/m$^2$K, approaching the ballistic limit in clean graphene proposed by Mingo and Brodio.[155] This temperature behavior of thermal conductance, $\propto T^{1.5}$, also agrees with the prediction that the out-of-plane acoustic phonon has dominant contribution to thermal conduction in suspended graphene.[155,156]

The thermal conductivity measured in the same work[154] is around 190 W/m-K at room temperature, which is one order of magnitude smaller than that from the Raman based measurements. This may be reasonable due to the differences in the sample length, as thermal conductivity in 2D lattice models has been predicted to be size dependent even before the discovery of graphene. Efforts have also been made to study the size dependent thermal conductivity in suspended CVD graphene based on Raman measurements.[150] However, the thermal conductivity was measured to randomly vary from ~2600 W/m-K to ~3100 W/m-K in corbino graphene with diameter range from 2.9 μm to 9.7 μm. The authors attributed this result to the uncertainty in Raman measurements, as well as the grain boundaries and defects in CVD graphene.

In addition, thermal conductivity in multilayer supported graphene was found to be size dependent.[143] The highest thermal conductivity measured is around 1250 W/m-K at room temperature in a 5 μm-long supported three layer graphene (Fig. 14 (a)). Interestingly, thermal conductivity decreases dramatically with decreased sample size; it reduces by 85% when the length of the sample reduces from 5 μm to 1 μm (S1 and S3 in Fig. 14). This size-dependent thermal conductivity is expected from the theory that $\kappa$ will keep increase with size even when the graphene flake is as long as 100 μm.[147]

Directly measuring the thermal contact resistance remains the primary challenge in studying the thermal conductivity in nanostructures, since it will contribute to the total measured thermal resistance. To investigate the intrinsic thermal conductivity of supported multilayer graphene, Wang *et al*. introduced a new method, noncontact electron beam heating technique, to measure the extrinsic thermal contact resistance directly.[143] As shown in Fig. 15 (a), multilayer graphene is supported by $SiN_x$ substrate with the two ends covered by Cr/Au thin films, labeled as B and D. An abrupt jump in the thermal resistance was observed at both B-C and C-D interfaces when the electron beam scanned across region B, C and D, indicating half of the total measured thermal resistance is from the two interfaces (shown in Fig. 15 (b)). This technique provides direct measurements of thermal contact resistance in nanostructures.

**IV. Application of Nanostructures for thermal management and renewable energy**

**A. Thermal Rectification in Nanostructures**

With the development of solid-state electronics devices, the integrated circuits can be built up by diodes and transistors. The conventional diodes and transistors can control the electrical conduction. In nature, besides electrons and photons, phonons can also transport and process information. It is straightforward that the realization of their counterpart of phononics device would have deep impacts. In recent years, more attention has been directed toward the phonon management on energy transport in dynamical systems and the emerging field is described as phononics.[28] Thermal diodes,[157-159] thermal transistors,[160] and thermal logic gates,[161] which are the basic components of functional thermal devices, have been proposed in theoretical model. Previously, the heat is regarded as useless or harmful in electronic circuits. The emerging of phononics raises the possibility that heat could be used to process information, in addition to electronics and photonics.

Like electronic counterpart, the thermal rectifier plays a vital role in phononics circuit. It is exciting that the first solid state thermal diode was fabricated in carbon nanotube junction [162] only two years after the theoretical prediction in two-segment lattice model.[158] In their experiment, Chang *et al*. modified the mass distribution and geometry (diameter) along the nanotubes by depositing amorphous $C_9H_{16}Pt$ at one end. The thermal rectification is measured to be around 2% in MWCNT, and 3%-7% in three different measured BNNTs, respectively.[162]

However, the rectifications in these CNT junctions are much lower than that of electronic counterparts. To enhance the rectifications, many simulation investigations have focused on various lattice modes[163-168] and asymmetric nanostructures,[169-171] like carbon nanocone(Fig. 16 (a)),[172,173] trapezia shaped GNRs (Fig. 16 (b)),[174] two rectangular GNRs with different widths (Fig. 16 (c)),[174] triangularly GNRs (Fig. 16 (d)),[175] and asymmetric three-terminal GNRs (Fig. 16 (e)).[176] The thermal rectification effect is observed in a Möbius graphene strip (Fig. 16 (f)),[115] which comes from the nonlinear interaction in graphene and the topological asymmetry of the Möobius strip. Interestingly, the rectification effect depends on the position of heat bath, since it can induce additional asymmetry. Moreover, it was found that the presence of layer-layer interactions in graphene Y junction (Fig. 16 (g))[110] may enhance the rectification effect. Thus multi-layer graphene is better than the single-layer counterpart in thermal rectification application.

Next we use carbon nanocone as an example to explore the thermal rectification and the underlying physical mechanism. As shown in Fig. 17 (c), the temperature at top/bottom of nanocone is set as $T_t = T_0(1-\Delta)$ /$T_b = T_0(1+\Delta)$, where $T_0$ is the average temperature, and $\Delta$ is the normalized temperature difference. To describe quantitatively the rectifier efficiency, we introduce thermal rectification ratio, $R \equiv 100\% \times (J_+ - J_-)/J_-$, where $J_+$ is the heat current from bottom to top and is the $J_-$ heat current from top to bottom.

Fig. 17 shows the thermal rectification effect in carbon nanocone. In Fig.17 (a), when thermal bias is positive, the heat flux $J_+$ increases steeply with $\Delta$. While in the region $\Delta < 0$, the heat flux $J_-$ is much smaller and changes a little with thermal bias. That is, this structure behaves as a "good" thermal conductor under positive "bias" and as a "poor" thermal conductor under negative "bias". The increase of $\Delta$ results in the increase of the rectification ratio. This is similar to the characteristic in electric diode. Besides homogeneous CNCs, CNCs with a graded mass distribution is also considered, in which the top atoms have minimum mass ($M_{C12}$) and the bottom atoms have maximum mass ($4M_{C12}$), where $M_{C12}$ is the mass of $^{12}$C atom. In CNTs with non-uniform axial heavy molecules deposit,[162] the atomic mass ratio is about 5 and the room temperature rectification is only 2% with $|\Delta|=0.05$. As shown in Fig. 17 (b), the rectification ratio is 10% for homogenous nanocone and 12% for graded massed nanocone with $\Delta=0.05$. This demonstrates that the geometric impact is more effective than the mass impact for high thermal rectification.

To understand the underlying mechanism of the rectification phenomenon, we show the phonon power spectra of atoms at the top and bottom atomic layers with temperature $T_0 = 300$ K and single-layers coupling in Fig. 18. The phonon power spectral density (PSD) describes the power carried by the phonon per unit frequency. A high PSD value for a phonon with frequency $f$ means that there are more states occupied by it. A zero means there is no such a phonon with $f$ exist in system. The phonon power spectrum analysis provides a noninvasive quantitative means of assessing the power carried by phonons in a system.

The spectra are obtained by Fourier transforming of velocities. Here, it is shown that the spectra along three directions, which were two in-plane directions and a out-plane direction. As shown in Fig. 18 (a), where the bottom atomic layer is at high temperature ($\Delta > 0$), the power spectra of the top and bottom atomic layers overlap perfectly in a large frequency range, which means the phonon can easily go through the nanocone along the direction of temperature gradient, and correspond to large $J_+$. On the contrary, when the top atomic layer contacted with high temperature bath ($\Delta < 0$, Fig. 18 (b)), there is an obvious mismatch in the spectra, both in low and high frequency band, which results in very low heat flux. The large mismatch in the spectra shows the weak correlation between the two ends. As a result, the phonon is difficult to go through the structure and thus leads to very small $J_-$. The match/mismatch of the power spectra between the bottom and top atomic layers controls the heat current and results in the rectification phenomenon.

In order to quantify the above power spectrum analysis, the overlaps (S) of the power spectra of the two layers are calculated as: $S_{\pm} = \int_0^{\infty} \min(P_{top}, P_{bot}) d\omega$, where $S_{+/-}$ corresponds to the case of $\Delta > 0$ and $\Delta < 0$, respectively. And $P_{top}$ is the power spectrum of the 4th (top) layer and $P_{bot}$ is the power spectrum of the 37th (bottom) layer. In Fig. 18, the value of $S_{+/-}$ is also shown. It illustrates that the heat flux and rectification phenomena correlate strongly with the overlap of the power spectra of the two ends.

Inspired by these theoretical and simulation studies, the solid state thermal rectifier were realized in various macroscopic systems,[177] and the rectification ratio was enhanced remarkably. Very recently, Tian *et al.* demonstrated the thermal rectification ratio up to 21% in trapezia shaped bulk graphene oxide.[178] Apart from the thermal rectifier, the acoustic rectifier was also realized,[179] which is fabricated by coupling a superlattice with a layer of ultrasound contrast agent microbubble suspension. A significant rectification with a rectifying ratio of ~ $10^4$ was observed within two frequency bands.

**B. Realization of Solid-State Thermal Memory**

Xie *et al.* have recently demonstrated a solid-state thermal memory which can process thermal information with two temperature states as input and output,[140] mimicking its electric counterpart. Their thermal memory device consists of three basic segments: an input terminal ($T_{in}$), an output terminal ($T_{out}$), and a heat conduction channel thermally bridging the two (Fig. 19 (a)). The two terminals, suspended $SiN_x$ membranes with Pt coil on top, are thermally connected to substrate ($T_{base}$) through six suspending leads, similar with that mentioned above. To study the thermal memory effect, Xie *et al.* used a single-crystalline suspended $VO_2$ nanobeam as a tunable thermal channel to obtain a nonlinear response between $T_{in}$ and $T_{out}$. A voltage bias, applied across the $VO_2$ nanobeam, is used to tune the characteristic of the thermal memory.

By applying forward and reverse temperature sweeping loop on $T_{in}$, Xie *et al.* presented a nonlinear and hysteresis $T_{in}$-$T_{out}$ curve, demonstrating a bi-stable temperature state (High/Low) of $T_{out}$ at the same input temperature $T_{in}$ (Fig. 19 (b)), which is an essence feature of a thermal memory. The hysteresis loop was enlarged by applying a bias voltage on $VO_2$ nanobeam and the operating temperature was shift downward (Fig. 19 (c)), which is also an important feature of a thermal memory. Both the High and Low temperature states at the output terminal are stable and repeatable over 150 cycles, with a slight fluctuation of ± 0.8K in the Low temperature state (Fig. 19 (d)). The thermal loop was proposed to be related to the first-order metal-insulator transition (MIT) in $VO_2$ nanobeam at around 340K, resulting in hysteresis in both electrical resistance and thermal conductance at the same temperature range.

**C. Enhancement of Thermoelectric Efficiency in Nano Structures**

Recently, thermoelectric materials have attracted extensive attention again. This is primarily due to the increasing awareness of the deleterious effect of global warming on the planet's environment, and a renewed requirement for long-life electrical power sources. Thermoelectric materials can provide electricity when subjected to a temperature gradient or provide cooling performance when electrical current is passed through it. For a good thermoelectric material, the material must have a high figure of merit (ZT), which is proportional to the Seebeck coefficient (S), electrical conductivity (σ), and absolute temperature, but inversely proportional to thermal conductivity.

Many thermoelectric materials have been identified till now, such as $Bi_2Te_3$, skutterudites $Co_4Sb_{12}$, SiGe alloys, PbTe, *et al.*.[180,181] High ZT can be achieved either by increasing the thermoelectric power-factor ($S^2\sigma$) or by decreasing thermal conductivity. Both methods can be realized via nanostructuring.

ZT enhancement has been achieved in bulk nanocomposites which consist of three-dimensional nanograins in traditional thermoelectric materials. Introducing nanostructures in bulk materials can contribute carriers to the host matrix, and they introduce variations in conduction/valence band edge which would make the effective band-structure different from that of the host matrix. Moreover, the very small size of the nano-grains introduces a high density of grain boundary interfaces. These numerous interfaces scatter phonons more strongly than charge carriers and decrease the phonon thermal conductivity while keeping thermoelectric power factor unchanged. Recently giant Seebeck coefficients have been achieved by controlling nanoparticles' size distribution.[182] And experiments show that the thermal conductivity decreases with grain size in the thermoelectric bulk materials. Thus by nanostructuring, enhancement in the power factor and reduction in the thermal conductivity are achievable simultaneously, resulting in significant enhancement in the value of ZT.[183-187]

Besides the bulk nanocomposites based traditional thermoelectric materials, it was found that low dimensional nanomaterials are promising candidates for solid state thermoelectric device applications although their bulk counterparts are with very low ZT, such as SiNWs. At room temperature, *ZT* of bulk silicon is only about 0.01. However, in SiNWs, the electrical conductivity and electron contribution to Seebeck coefficient are similar to those of bulk silicon, but exhibit 100-fold reduction in thermal conductivity, showing that an approximately 100-fold improvement of the ZT values over bulk Si are achieved in SiNW.[134,188] Particularly, theories and experiments indicated that a larger reduction in thermal conductivity can be achieved in nanometer-sized low-dimensional structures, due to the boundary and interface phonon-scattering mechanisms.[189,190] It was proposed to use quantum size effects in low-dimensional materials to create sharp features in the density-of-states and enhance the Seebeck coefficient.[191-193] Benefit from the reduced thermal conductivity and/or increased power factor, enhanced thermoelectric performance have been observed in 1D NWs,[194-197] NTs[55,198] and quasi-2D nanoribbons.[97,117,199-202]

**V. Conclusion and Outlook**

In this paper we present the state-of-the-art of topic about thermal transport in nanoscale materials. We firstly reviewed the fundamental physical phenomena for thermal transport in low dimensional systems, and various impacts on thermal conductivity of nanotubes, nanowires and graphenes. Then, we provide a review on the recent experimental measurements on thermal conductivity and temperature in nanoscale. In these low dimensional structures, phonon transports super-diffusively, which leads to a length dependent thermal conductivity. Thus nanoscale materials are promising platforms to verify fundamental thermal transport theories. Moreover, current status of research on applications of nanomaterials, include renewable energy and thermal management in nanoscale, has been reviewed in this paper.

Based on the present state of the art reviewed here, a significant progress on understanding of thermal transport in nanoscale was achieved in the past decade. However, there are still many questions and open challenges remaining. On the experimental side, one challenge is how to make the sample smaller, towards a dozens of nanometers or even a few nanometers only. Another challenge lies in the difficulties in measuring the thermal contact resistance at the two ends of sample. On the theoretical side, it is necessary to establish an improved theory describing thermal transport across the interface by taking into account the anomalous thermal transport characteristics of nanostructures. In addition, how to set up a transport theory by incorporating nonlinearity in the quantum regime is still a challenge. All these deserve further systematical investigations from a joint experimental and theoretical effort.

**Acknowledgements**

This work was supported in part by the National Natural Science Foundation of China (Grant No. 11204216) (NY) and the startup fund from Tongji University (NY, XX and BL), from National University of Singapore (W-144-000-285-646), and the National Natural Science Foundation of China (Grant No. 11274011) (GZ), the Ministry of Education of China (Grant No. 20110001120133) (GZ), and the Ministry of Science and Technology of China (Grant No. 2011CB933001) (GZ).

References

[1] S. Maruyama, Microscale Therm. Eng. **7**, 41 (2003).

[2] S.G. Volz and G. Chen, Appl. Phys. Lett. **75**, 2056 (1999).

[3] A.A. Balandin, S. Ghosh, W. Bao, I. Calizo, D. Teweldebrhan, F. Miao, and C.N. Lau, Nano Lett. **8**, 902 (2008).

[4] G. Chen, J. Heat Transf. **119**, 220 (1997).

[5] K. Huang and B.-F. Zhu, Phys. Rev. B **38**, 2183 (1988).

[6] B. Li and J. Wang, Phys. Rev. Lett. **91**, 044301 (2003).

[7] J.-S. Wang and B. Li, Phys. Rev. Lett. **92**, 074302 (2004).

[8] H. Zhao, L. Yi, F. Liu, and B. Xu, Eur. Phys. J. B **54**, 185 (2006).

[9] O. Narayan and S. Ramaswamy, Phys. Rev. Lett. **89**, 200601 (2002).

[10] A. Dhar, Adv. Phys. **57**, 457 (2008).

[11] S. Lepri, R. Livi, and A. Politi, Phys. Rev. E **68**, 067102 (2003).

[12] G. Chen, Phys. Rev. Lett. **86**, 2297 (2001).

[13] N. Li, P. Tong, and B. Li, Europhys. Lett. **75**, 49 (2006).

[14] F.X. Alvarez, D. Jou, and A. Sellitto, J. Appl. Phys. **105**, 014317 (2009).

[15] F.X. Alvarez, D. Jou, and A. Sellitto, J. Heat Transf. **133**, 022402 (2011).

[16] D.Y. Tzou and Z.-Y. Guo, Inter. J.Therm. Sci. **49**, 1133 (2010).

[17] N. Yang, G. Zhang, and B. Li, Nano Today **5**, 85 (2010).

[18] J. Chen, G. Zhang, and B. Li, Phys. Soc. JPN **79**, 074604 (2010).

[19] J. Chen, G. Zhang, and B. Li, Phys. Lett. A **374**, 2392 (2010).

[20] B.-Y. Cao and Y.-W. Li, J. Chem. Phys. **133**, 024106 (2010).

[21] B.-Y. Cao, J. Chem. Phys. **129**, 074106 (2008).

[22] Z. Wang, R. Zhao, and Y. Chen, Sci. China Ser. E **53**, 429 (2010).

[23] N. Yang, X. Ni, J.-W. Jiang, and B. Li, Appl. Phys. Lett. **100**, 093107 (2012).

[24] Y. Xu, J.-S. Wang, W. Duan, B.-L. Gu, and B. Li, Phys. Rev. B **78**, 224303 (2008).

[25] G. Chen, J. Heat Transf. **121**, 945 (1999).

[26] D.G. Cahill, W.K. Ford, K.E. Goodson, G.D. Mahan, A. Majumdar, H.J. Maris, R. Merlin, and S.R. Phillpot, J. Appl. Phys. **93**, 793 (2003).

[27] B. Li, J. Wang, L. Wang, and G. Zhang, Chaos **15**, 015121 (2005).

[28] N. Li, J. Ren, L. Wang, G. Zhang, P. Hänggi, and B. Li, Rev. Mod. Phys. **84**, 1045 (2012).

[29] G. Zhang and B. Li, Nanoscale **2**, 1058 (2010).

[30] E. Pop, Nano Research **3**, 147 (2010).

[31] Y. Dubi and M. Di Ventra, Rev. Mod. Phys. **83**, 131 (2011).

[32] S. Liu, X. Xu, R. Xie, G. Zhang, and B. Li, Eur. Phys. J. B **85**, 1 (2012).

[33] N. Yang, "Thermal Transport in Low Dimensional Graded Structures and Silicon Nanowires", Ph.D Thesis, National University of Singapore, Singapore (2009).

[34] J. Chen, "Theoretical Investigation on Thermal Properties of Silicon Based Nanostructures", Ph.D Thesis, National University of Singapore, Singapore (2011).

[35] L. Shi, "Electrical-thermal energy transfer and energy conversion in semiconductor nanowires", Ph.D Thesis, National University of Singapore, Singapore (2011).

[36] S. Iijima, Nature **354**, 56 (1991).

[37] S. Berber, Y.-K. Kwon, and D. Tománek, Phys. Rev. Lett. **84**, 4613 (2000).

[38] P. Kim, L. Shi, A. Majumdar, and P.L. McEuen, Phys. Rev. Lett. **87**, 215502 (2001).

[39] G. Zhang and B. Li, J. Phys. Chem. B **109**, 23823 (2005).

[40] D.J. Yang, Q. Zhang, G. Chen, S.F. Yoon, J. Ahn, S.G. Wang, Q. Zhou, Q. Wang, and J.Q. Li, Phys. Rev. B **66**, 165440 (2002).

[41] C.W. Padgett and D.W. Brenner, Nano Lett. **4**, 1051 (2004).

[42] J.F. Moreland, Microscale Therm. Eng. **8**, 61 (2004).

[43] S. Maruyama, Physica B **323**, 193 (2002).

[44] C.W. Chang, D. Okawa, H. Garcia, A. Majumdar, and A. Zettl, Phys. Rev. Lett. **101**, 075903 (2008).

[45] G. Zhang and B. Li, J. Chem. Phys. **123**, 114714 (2005).

[46] Z.-X. Guo and X.-G. Gong, Fron. Phys. China **4**, 389 (2009).

[47] H. Zhu, Y. Xu, B.-L. Gu, and W. Duan, New J. Phys. **14**, 013053/1 (2012).

[48] Z.-Y. Ong and E. Pop, Phys. Rev. B **81**, 155408 (2010).

[49] Z. Guo, D. Zhang, Y. Zhai, and X.-G. Gong, Nanotechnology **21**, 285706 (2010).

[50] Z.-X. Guo, D. Zhang, and X.-G. Gong, Phys. Rev. B **84**, 075470 (2011).

[51] W. Lin, J. Shang, W. Gu, and C.P. Wong, Carbon **50**, 1591 (2012).

[52] Y. Gu and Y. Chen, Phys. Rev. B **76**, 134110 (2007).

[53] J. Yang, Y. Yang, S.W. Waltermire, T. Gutu, A.A. Zinn, T.T. Xu, Y. Chen, and D. Li, Small **7**, 2334 (2011).

[54] J. Yang, S. Waltermire, Y. Chen, A.A. Zinn, T.T. Xu, and D. Li, Appl. Phys. Lett. **96**, 023109 (2010).

[55] X. Tan, H. Liu, Y. Wen, H. Lv, L. Pan, J. Shi, and X. Tang, Nanaoscale. Res. Lett. **7**, 116 (2012).

[56] Y.-F. Gao, Q.-Y. Meng, L. Zhang, J.-Q. Liu, and Y.-H. Jing, Acta. Phys.-Chim. Sin. **28**, 1077.

[57] J. Wang, L. Li, and J.-S. Wang, Appl. Phys. Lett. **99**, 091905 (2011).

[58] C. Ren, W. Zhang, Z. Xu, Z. Zhu, and P. Huai, J. Phys. Chem. C **114**, 5786 (2010).

[59] C. Ren, Z. Xu, W. Zhang, Y. Li, Z. Zhu, and P. Huai, Phys. Lett. A **374**, 1860 (2010).

[60] X. Wang, Z. Huang, T. Wang, Y.W. Tang, and X.C. Zeng, Physica B **403**, 2021 (2008).

[61] P. Rui-Qin, X. Zi-Jian, and Z. Zhi-Yuan, Chinese Phys. Lett. **24**, 1321 (2007).

[62] T. Yamamoto and K. Watanabe, Phys. Rev. Lett. **96**, 255503 (2006).

[63] C.W. Chang, A.M. Fennimore, A. Afanasiev, D. Okawa, T. Ikuno, H. Garcia, D. Li, A. Majumdar, and A. Zettl, Phys. Rev. Lett. **97**, 089501 (2006).

[64] G. Zhang and B. Li, J. Chem. Phys. **123**, 014705 (2005).

[65] Y. Cui, Q. Wei, H. Park, and C.M. Lieber, Science **293**, 1289 (2001).

[66] G.-J. Zhang, G. Zhang, J.H. Chua, R.-E. Chee, E.H. Wong, A. Agarwal, K.D. Buddharaju, N. Singh, Z. Gao, and N. Balasubramanian, Nano Lett. **8**, 1066 (2008).

[67] J. Xiang, W. Lu, Y. Hu, Y. Wu, H. Yan, and C.M. Lieber, Nature **441**, 489 (2006).

[68] S.C. Rustagi, N. Singh, Y.F. Lim, G. Zhang, S. Wang, G.Q. Lo, N. Balasubramanian, and D.L. Kwong, IEEE Electr. Device L. **28**, 909 (2007).

[69] L. Hu and G. Chen, Nano Lett. **7**, 3249 (2007).

[70] J. Li, H. Yu, S.M. Wong, G. Zhang, X. Sun, P.G.-Q. Lo, and D.-L. Kwong, Appl. Phys. Lett. **95**, 033102 (2009).

[71] D. Li, Y. Wu, P. Kim, L. Shi, P. Yang, and A. Majumdar, Appl. Phys. Lett. **83**, 2934 (2003).

[72] Y. Chen, D. Li, J.R. Lukes, and A. Majumdar, J. Heat Transf. **127**, 1129 (2005).

[73] D. Yao, G. Zhang, and B. Li, Nano Lett. **8**, 4557 (2008).

[74] D. Yao, G. Zhang, G.-Q. Lo, and B. Li, Appl. Phys. Lett. **94**, 113113 (2009).

[75] J. Chen, G. Zhang, and B. Li, J. Chem. Phys. **135**, 104508 (2011).

[76] J. Chen, G. Zhang, and B. Li, J. Chem. Phys. **135**, 204705 (2011).

[77] D. Donadio and G. Galli, Phys. Rev. Lett. **102**, 195901 (2009).

[78] Y. Chen, D. Li, J. Yang, Y. Wu, J.R. Lukes, and A. Majumdar, Physica B **349**, 270 (2004).

[79] N. Yang, G. Zhang, and B. Li, Nano Lett. **8**, 276 (2008).

[80] P.K. Schelling, S.R. Phillpot, and P. Keblinski, Phys. Rev. B **65**, 144306 (2002).

[81] L.H. Liang and B. Li, Phys. Rev. B **73**, 153303 (2006).

[82] M. Wang, N. Yang, and Z.-Y. Guo, J. Appl. Phys. **110**, 064310 (2011).

[83] Z.-X. Xie, K.-Q. Chen, and L.-M. Tang, J. Appl. Phys. **110**, 124321 (2011).

[84] X.-F. Peng, K.-Q. Chen, Q. Wan, B.S. Zou, and W. Duan, Phys. Rev. B **81**, 195317 (2010).

[85] X.-F. Peng and K.-Q. Chen, Physica E **42**, 1968 (2010).

[86] X.-F. Peng and X.-J. Wang, J. Appl. Phys. **110**, 044305 (2011).

[87] J. Chen, G. Zhang, and B. Li, Appl. Phys. Lett. **95**, 073117 (2009).

[88] J. Chen, G. Zhang, and B. Li, Nano Lett. **10**, 3978 (2010).

[89] H.P. Li, A.D. Sarkar, and R.Q. Zhang, Europhys. Lett. **96**, 56007 (2011).

[90] G. Chen, Phys. Rev. B **57**, 14958 (1998).

[91] J. Chen, G. Zhang, and B. Li, Nano Lett. **12**, 2826 (2012).

[92] M.C. Wingert, Z.C.Y. Chen, E. Dechaumphai, J. Moon, J.-H. Kim, J. Xiang, and R. Chen, Nano Lett. **11**, 5507 (2011).

[93] K.S. Novoselov, A.K. Geim, S.V. Morozov, D. Jiang, Y. Zhang, S.V. Dubonos, I.V. Grigorieva, and A.A. Firsov, Science **306**, 666 (2004).

[94] Y. Zhang, Y.-W. Tan, H.L. Stormer, and P. Kim, Nature **438**, 201 (2005).

[95] D. Xiong, J. Wang, Y. Zhang, and H. Zhao, Phys. Rev. E **82**, 030101 (2010).

[96] W.J. Evans, L. Hu, and P. Keblinski, Appl. Phys. Lett. **96**, 203112 (2010).

[97] H. Zheng, H.J. Liu, X.J. Tan, H.Y. Lv, L. Pan, J. Shi, and X.F. Tang, Appl. Phys. Lett. **100**, 093104 (2012).

[98] J. Zhang, X. Huang, Y. Yue, J. Wang, and X. Wang, Phys. Rev. B **84**, 235416 (2011).

[99] Z.-X. Xie, K.-Q. Chen, and W. Duan, J. Phys. Condens. Mat. **23**, 315302 (2011).

[100] J.-W. Jiang, J.-S. Wang, and B. Li, Phys. Rev. B **79**, 205418/1 (2009).

[101] J.W. Jiang, J.H. Lan, J.S. Wang, and B.W. Li, J. Appl. Phys. **107**, 054314 (2010).

[102] T. Ouyang, Y.P. Chen, K.K. Yang, and J.X. Zhong, Europhys. Lett. **88**, 28002 (2009).

[103] X.-F. Peng, X.-J. Wang, Z.-Q. Gong, and K.-Q. Chen, Appl. Phys. Lett. **99**, 233105 (2011).

[104] X.-F. Peng, X.-J. Wang, L.-Q. Chen, and K.-Q. Chen, Europhys. Lett. **98**, 56001 (2012).

[105] X. Li, K. Maute, M.L. Dunn, and R. Yang, Phys. Rev. B **81**, 245318 (2010).

[106] X. Zhai and G. Jin, Europhys. Lett. **96**, 16002 (2011).

[107] N. Wei, L.Q. Xu, H.Q. Wang, and J.C. Zheng, Nanotechnology **22** (2011).

[108] J.-W. Jiang, J.-S. Wang, and B. Li, Phys. Rev. B **80**, 113405/1 (2009).

[109] Z.-X. Guo, J.W. Ding, and X.-G. Gong, Phys. Rev. B **85**, 235429 (2012).

[110] G. Zhang and H. Zhang, Nanoscale **3**, 4604 (2011).

[111] W.-R. Zhong, M.-P. Zhang, B.-Q. Ai, and D.-Q. Zheng, Appl. Phys. Lett. **98**, 113107 (2011).

[112] H.-Y. Cao, Z.-X. Guo, H. Xiang, and X.-G. Gong, Phys. Lett. A **376**, 525 (2012).

[113] Y. Xu, X. Chen, J.-S. Wang, B.-L. Gu, and W. Duan, Phys. Rev. B **81**, 195425 (2010).

[114] Y. Xu, X. Chen, B.-L. Gu, and W. Duan, Appl. Phys. Lett. **95**, 233116 (2009).

[115] J.-W. Jiang, J.-S. Wang, and B. Li, J. Appl. Phys. **108**, 064307/1 (2010).

[116] J.-W. Jiang and J.-S. Wang, Phys. Rev. B **81**, 174117 (2010).

[117] Z.-X. Xie, L.-M. Tang, C.-N. Pan, K.-M. Li, K.-Q. Chen, and W. Duan, Appl. Phys. Lett. **100**, 073105 (2012).

[118] Z.-X. Guo, D. Zhang, and X.-G. Gong, Appl. Phys. Lett. **95**, 163103 (2009).

[119] T. Ouyang, Y. Chen, L.-M. Liu, Y. Xie, X. Wei, and J. Zhong, Phys. Rev. B **85**, 235436 (2012).

[120] T. Ouyang, Y. Chen, Y. Xie, G.M. Stocks, and J. Zhong, Appl. Phys. Lett. **99**, 233101 (2011).

[121] Z. Wei, Z. Ni, K. Bi, M. Chen, and Y. Chen, Phys. Lett. A **375**, 1195 (2011).

[122] Z. Wei, Z. Ni, K. Bi, M. Chen, and Y. Chen, Carbon **49**, 2653 (2011).

[123] Z.-G. Bao, Y.-P. Chen, T. Ouyang, and e. al, Acta Phys. Sin. **60**, 028103 (2011).

[124] T. Ouyang, Y. Chen, Y. Xie, K. Yang, Z. Bao, and J. Zhong, Nanotechnology **21**, 245701 (2010).

[125] K. Yang, Y. Chen, Y. Xie, X.L. Wei, T. Ouyang, and J. Zhong, Solid State Commun. **151**, 460 (2011).

[126] B.-Q. Ai, W.-R. Zhong, and B. Hu, J. Phys. Chem. C **116**, 13810 (2012).

[127] R. Ma, L. Zhu, L. Sheng, M. Liu, and D.N. Sheng, Phys. Rev. B **84**, 075420 (2011).

[128] F. Hao, D. Fang, and Z. Xu, Appl. Phys. Lett. **99**, 041901 (2011).

[129] K. Yang, Y. Chen, Y. Xie, T. Ouyang, and J. Zhong, Europhys. Lett. **91**, 46006 (2010).

[130] L. Lu, W. Yi, and D.L. Zhang, Rev. Sci. Instrum. **72**, 2296 (2001).

[131] J. Hone, M. Whitney, C. Piskoti, and A. Zettl, Phys. Rev. B **59**, R2514 (1999).

[132] W. Yi, L. Lu, D.L. Zhang, Z.W. Pan, and S.S. Xie, Phys. Rev. B **59**, R9015 (1999).

[133] R. Chen, A.I. Hochbaum, P. Murphy, J. Moore, P. Yang, and A. Majumdar, Phys. Rev. Lett. **101**, 105501 (2008).

[134] A.I. Hochbaum, R. Chen, R.D. Delgado, W. Liang, E.C. Garnett, M. Najarian, A. Majumdar, and P. Yang, Nature **451**, 163 (2008).

[135] K. Hippalgaonkar, B. Huang, R. Chen, K. Sawyer, P. Ercius, and A. Majumdar, Nano Lett. **10**, 4341 (2010).

[136] L. Shi, D. Li, C. Yu, W. Jang, D. Kim, Z. Yao, P. KIM, and A. Majumdar, J. Heat Transf. **125**, 881 (2003).

[137] C.W. Chang, Science **320**, 1121 (2008).

[138] J. Yang, Y. Yang, S. Waltermire, X. Wu, H. Zhang, T. Gutu, Y. Jiang, Y. Chen, A. Zinn, R. Prasher, T. Xu, and D. Li, Nat. Nanotech. **7**, 91 (2011).

[139] C. Bui, R. Xie, M. Zheng, Q. Zhang, C. Sow, B. Li, and J. Thong, Small **8**, 738 (2012).

[140] R. Xie, C. Bui, B. Varghese, Q. Zhang, C. Sow, B. Li, and J. Thong, Adv. Funct. Mater. **21**, 1602 (2011).

[141] M. Wingert, Z. Chen, E. Dechaumphai, J. Moon, J.-H. Kim, J. Xiang, and R. Chen, Nano Lett. **11**, 5507 (2011).

[142] M.C. Wingert, Z.C.Y. Chen, S. Kwon, J. Xiang, and R. Chen, Rev. Sci. Instrum. **83**, 024901 (2012).

[143] Z. Wang, R. Xie, C. Bui, D. Liu, X. Ni, B. Li, and J. Thong, Nano Lett. **11**, 113 (2011).

[144] J. Yang, Y. Yang, S.W. Waltermire, X. Wu, H. Zhang, T. Gutu, Y. Jiang, Y. Chen, A.A. Zinn, R. Prasher, T.T. Xu, and D. Li, Nat Nano **7**, 91 (2012).

[145] K.S. Novoselov, A.K. Geim, S.V. Morozov, D. Jiang, M.I. Katsnelson, I.V. Grigorieva, S.V. Dubonos, and A.A. Firsov, Nature **438**, 197 (2005).

[146] S. Ghosh, I. Calizo, D. Teweldebrhan, E.P. Pokatilov, D.L. Nika, A.A. Balandin, W. Bao, F. Miao, and C.N. Lau, Appl. Phys. Lett. **92**, 151911 (2008).

[147] D.L. Nika, S. Ghosh, E.P. Pokatilov, and A.A. Balandin, Appl. Phys. Lett. **94**, 203103 (2009).

[148]S. Ghosh, W. Bao, D.L. Nika, S. Subrina, E.P. Pokatilov, C.N. Lau, and A.A. Balandin, Nat. Mater. **9**, 555 (2010).

[149]W. Cai, A.L. Moore, Y. Zhu, X. Li, S. Chen, L. Shi, and R.S. Ruoff, Nano Lett. **10**, 1645 (2010).

[150]S. Chen, A.L. Moore, W.W. Cai, J. Suk, J. An, C. Mishra, C. Amos, A.W. Magnuson, J.Y. Kang, L. Shi, and R.S. Ruoff, ACS Nano **5**, 321 (2011).

[151]J.-U. Lee, D. Yoon, H. Kim, S.W. Lee, and H. Cheong, Phys. Rev. B **83**, 081419 (2011).

[152]J.H. Seol, I. Jo, A.L. Moore, L. Lindsay, Z.H. Aitken, M.T. Pettes, X. Li, Z. Yao, R. Huang, D. Broido, N. Mingo, R.S. Ruoff, and L. Shi, Science **328**, 213 (2010).

[153]S. Chen, Q. Wu, C. Mishra, J. Kang, H. Zhang, K. Cho, W. Cai, A.A. Balandin, and R.S. Ruoff, Nat. Mater. **11**, 203 (2012).

[154]X. Xu, Y. Wang, K. Zhang, X. Zhao, S. Bae, M. Heinrich, C.T. Bui, R. Xie, J.T.L. Thong, B.H. Hong, K.P. Loh, B. Li, and B. Oezyilmaz, arXiv:1012.2937 (2010).

[155]N. Mingo and D.A. Broido, Phys. Rev. Lett. **95**, 096105 (2005).

[156]E. Muñoz, J. Lu, and B.I. Yakobson, Nano Lett. **10**, 1652 (2010).

[157]M. Terraneo, M. Peyrard, and G. Casati, Phys. Rev. Lett. **88**, 094302 (2002).

[158]B. Li, L. Wang, and G. Casati, Phys. Rev. Lett. **93**, 184301 (2004).

[159]N. Yang, N. Li, L. Wang, and B. Li, Phys. Rev. B **76**, 020301 (2007).

[160]B. Li, L. Wang, and G. Casati, Appl. Phys. Lett. **88**, 143501 (2006).

[161]L. Wang and B. Li, Phys. Rev. Lett. **99**, 177208 (2007).

[162]C.W. Chang, D. Okawa, A. Majumdar, and A. Zettl, Science **314**, 1121 (2006).

[163]Y. Yan, Q.-F. Liang, and H. Zhao, Phys. Lett. A **375**, 4074 (2011).

[164]Y. Yan and H. Zhao, J. Phys. Condens. Mat. **24**, 275401 (2012).

[165]W.-R. Zhong, P. Yang, B.-Q. Ai, Z.-G. Shao, and B. Hu, Phys. Rev. E **79**, 050103 (2009).

[166]B.-Q. Ai, W.-R. Zhong, and B. Hu, Phys. Rev. E **83**, 052102 (2011).

[167]B.-Q. Ai and B. Hu, Phys. Rev. E **83**, 011131 (2011).

[168]B.-Q. Ai, D. He, and B. Hu, Phys. Rev. E **81**, 031124 (2010).

[169]S.-C. Wang and X.-G. Liang, Inter. J.Therm. Sci. **50**, 680 (2011).

[170]W.-R. Zhong, W.-H. Huang, X.-R. Deng, and B.-Q. Ai, Appl. Phys. Lett. **99**, 193104 (2011).

[171]G. Wu and B. Li, Phys. Rev. B **76**, 085424 (2007).

[172]N. Yang, G. Zhang, and B. Li, Appl. Phys. Lett. **93**, 243111 (2008).

[173]G. Wu and B. Li, J. Phys. Condens. Mat. **20**, 175211 (2008).

[174]N. Yang, G. Zhang, and B. Li, Appl. Phys. Lett. **95**, 033107 (2009).

[175]J. Hu, X. Ruan, and Y.P. Chen, Nano Lett. **9**, 2730 (2009).

[176]T. Ouyang, Y. Chen, Y. Xie, X.L. Wei, K. Yang, P. Yang, and J. Zhong, Phys. Rev. B **82**, 245403 (2010).

[177]W. Kobayashi, Y. Teraoka, and I. Terasaki, Appl. Phys. Lett. **95**, 171905 (2009).

[178]H. Tian, D. Xie, Y. Yang, T.-L. Ren, G. Zhang, Y.-F. Wang, C.-J. Zhou, P.-G. Peng, L.-G. Wang, and L.-T. Liu, Sci. Rep. **2** (2012).

[179]B. Liang, X.S. Guo, J. Tu, D. Zhang, and J.C. Cheng, Nat. Mater. **9**, 989 (2012).

[180]Y. Lan, A.J. Minnich, G. Chen, and Z. Ren, Adv. Funct. Mater. **20**, 357 (2010).

[181]C.J. Vineis, A. Shakouri, A. Majumdar, and M.G. Kanatzidis, Advanced Materials **22**, 3970 (2010).

[182]M. Zebarjadi, K. Esfarjani, Z. Bian, and A. Shakouri, Nano Lett. **11**, 225 (2010).

[183]W. Xie, J. He, H.J. Kang, X. Tang, S. Zhu, M. Laver, S. Wang, J.R.D. Copley, C.M. Brown, Q. Zhang, and T.M. Tritt, Nano Lett. **10**, 3283 (2010).

[184]X.J. Tan, W. Liu, H.J. Liu, J. Shi, X.F. Tang, and C. Uher, Phys. Rev. B **85**, 205212 (2012).

[185]W. Liu, X. Tang, H. Li, K. Yin, J. Sharp, X. Zhou, and C. Uher, J. Mater. Chem. **22**, 13653 (2012).

[186]W. Liu, X. Tan, K. Yin, H. Liu, X. Tang, J. Shi, Q. Zhang, and C. Uher, Phys. Rev. Lett. **108**, 166601 (2012).

[187]X. Su, H. Li, Y. Yan, G. Wang, H. Chi, X. Zhou, X. Tang, Q. Zhang, and U. Ctirad, Acta Mater. **60**, 3536 (2012).

[188]A.I. Boukai, Y. Bunimovich, J. Tahir-Kheli, J.K. Yu, W.A. Goddard, and J.R. Heath, Nature **451**, 168 (2008).

[189]Y. Chen, D. Li, J. Yang, Y. Wang, and H. Hu, J. Comput. Theor. NanoS. **5**, 157 (2008).

[190]Y. Chen, D. Li, J.R. Lukes, Z. Ni, and M. Chen, Phys. Rev. B **72**, 174302 (2005).

[191]L.D. Hicks and M.S. Dresselhaus, Phys. Rev. B **47**, 16631 (1993).

[192]L. Shi, D. Yao, G. Zhang, and B. Li, Appl. Phys. Lett. **95**, 063102 (2009).

[193]L. Shi, D. Yao, G. Zhang, and B. Li, Appl. Phys. Lett. **96**, 173108 (2010).

[194]G. Zhang, Q. Zhang, C.-T. Bui, G.-Q. Lo, and B. Li, Appl. Phys. Lett. **94**, 213108 (2009).

[195]G. Zhang, Q.-X. Zhang, D. Kavitha, and G.-Q. Lo, Appl. Phys. Lett. **95**, 243104 (2009).

[196]L. Shi, J. Chen, G. Zhang, and B. Li, Phys. Lett. A **376**, 978 (2012).

[197]C. Liu and J. Li, Phys. Lett. A **375**, 2878 (2011).

[198]H.G. Si, Y.X. Wang, Y.L. Yan, and G.B. Zhang, J. Phys. Chem. C **116**, 3956 (2012).

[199]H.Y. Lv, H.J. Liu, X.J. Tan, L. Pan, Y.W. Wen, J. Shi, and X.F. Tang, Nanoscale **4**, 511 (2012).

[200]K. Yang, Y. Chen, R. D'Agosta, Y. Xie, J. Zhong, and A. Rubio, Phys. Rev. B **86**, 045425 (2012).

[201]X. Chen, Z. Wang, and Y. Ma, J. Phys. Chem. C **115**, 20696 (2011).

[202]X. Ni, G. Liang, J.-S. Wang, and B. Li, Appl. Phys. Lett. **95**, 192114 (2009).

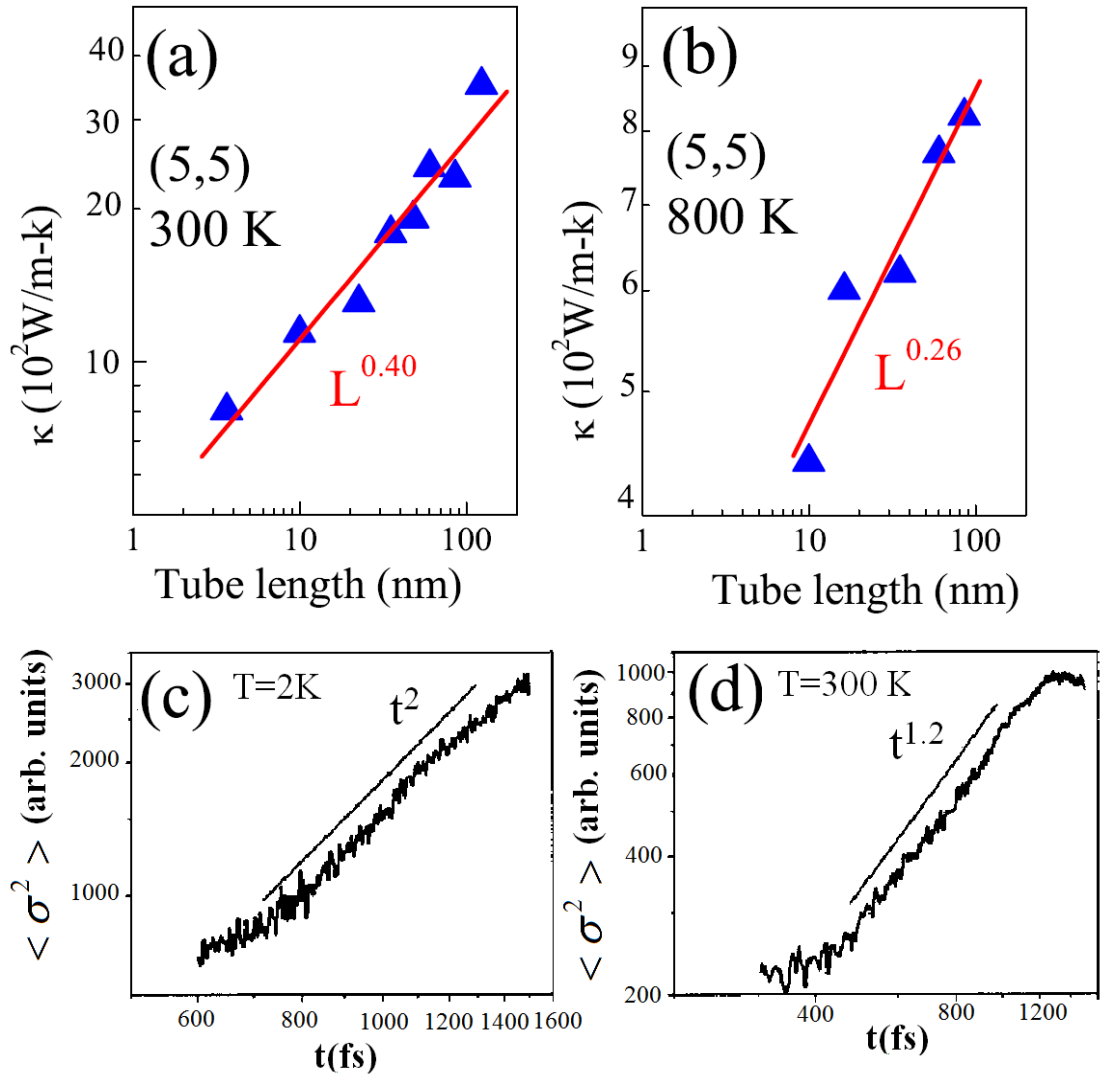

Fig.1 (Color online) (a) and (b)The thermal conductivity versus carbon nanotube length in log-log scale for (5,5) SWCNT at 300K and 800K. The thermal conductivity of carbon nanotube diverged to the length, as $\kappa \propto L^{\beta}$. (c) and (d) Energy diffusion in a SWCNT at 2K and room temperature.

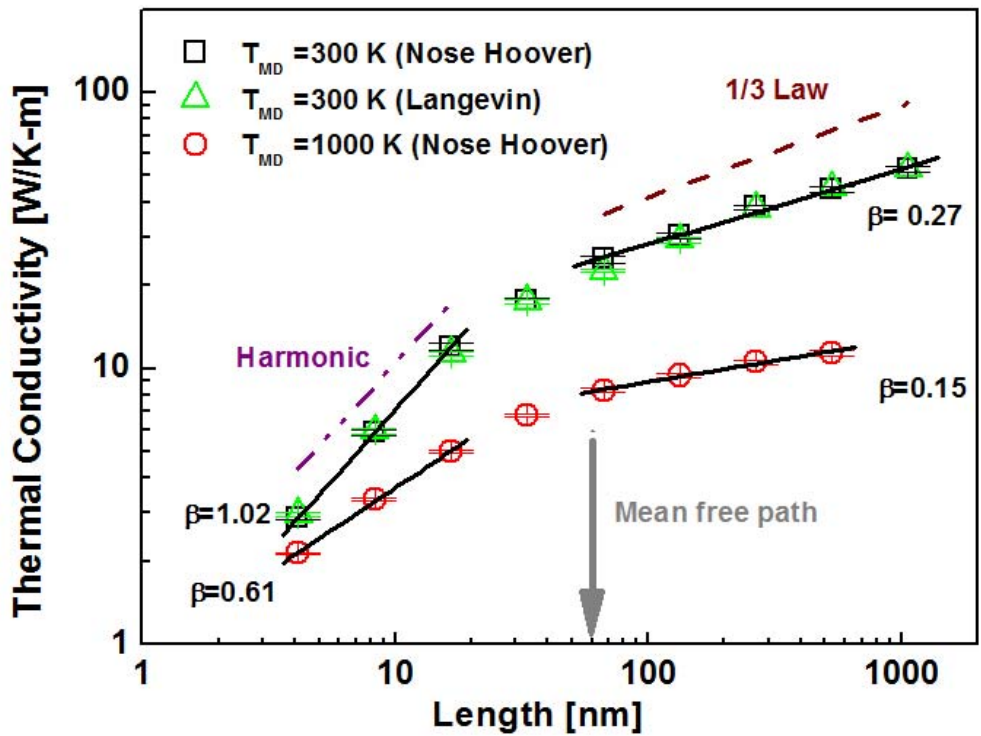

Fig.2 (Color online) The thermal conductivity of SiNWs (with fixed transverse boundary condition) vs length. The black solid lines are the best fitting ones. The harmonic (dash dotted) and 1/3 (dashed) laws are shown for reference.

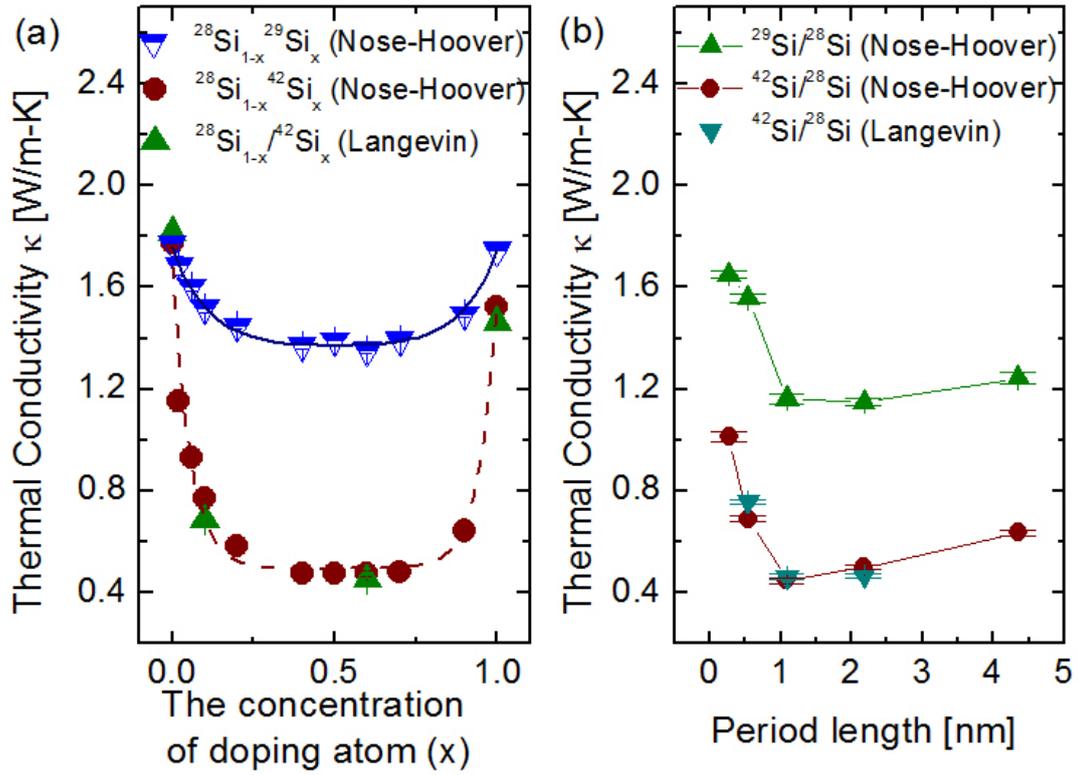

Fig.3 (Color online) (a) Thermal conductivity of SiNWs versus the percentage of randomly doping isotope atoms at 300K. The results by Nose-Hoover method coincide with those by Langevin methods indicating that the results are independent of the heat bath used. The solid curve and the dash curve are the best fitting to the formula $\kappa = A_1 e^{-x/B} + A_2 e^{-(1-x)/B} + C$. (b) Thermal conductivity of the superlattice SiNWs versus the period length at 300 K.

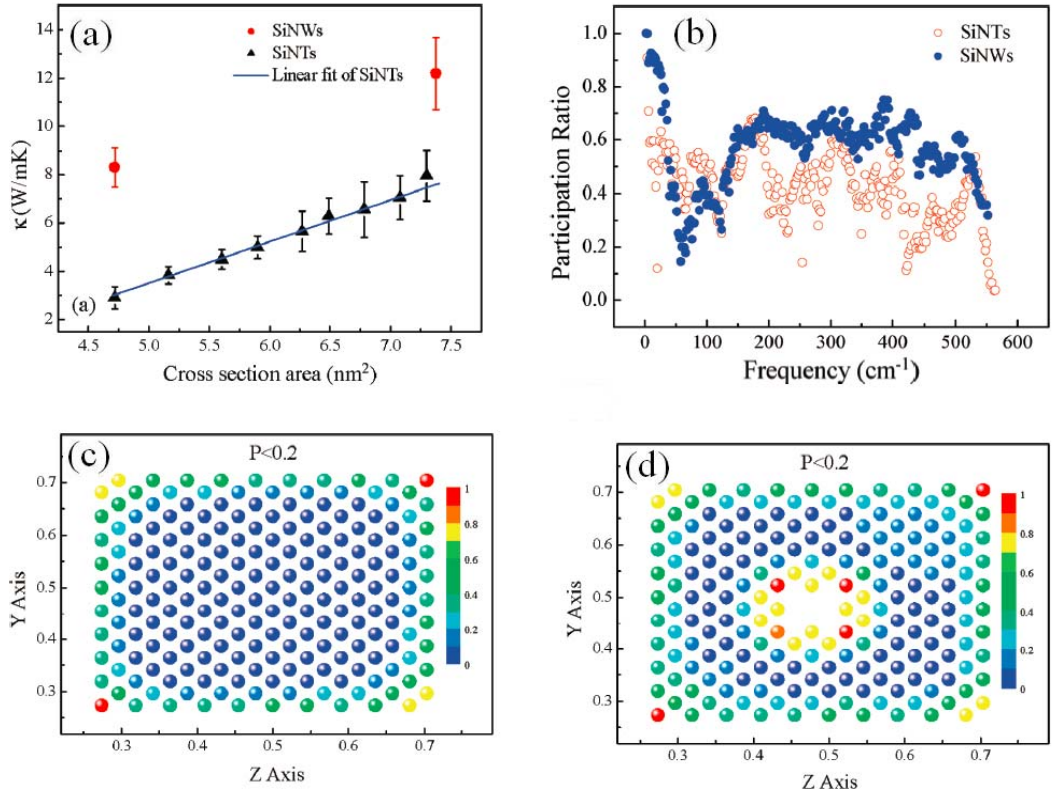

Fig.4 (Color online) (a) Room temperature thermal conductivity of SiNWs (read) and SiNTs (black) versus cross section area. (b) P-ratio of each eigen-mode for SiNTs (red) and SiNWs (blue) with the same cross section area. The p-ratio measures the fraction of atoms participating in a given mode, and effectively indicates the localized modes with O(1/N) and delocalized modes with O(1). (c) and (d) are normalized energy distribution on the cross section for SiNWs and SiNTs at 300 K, respectively. Positions of the circles denote the different locations on the plane, and intensity of the energy is depicted according to the color bar. P is participation ratio.

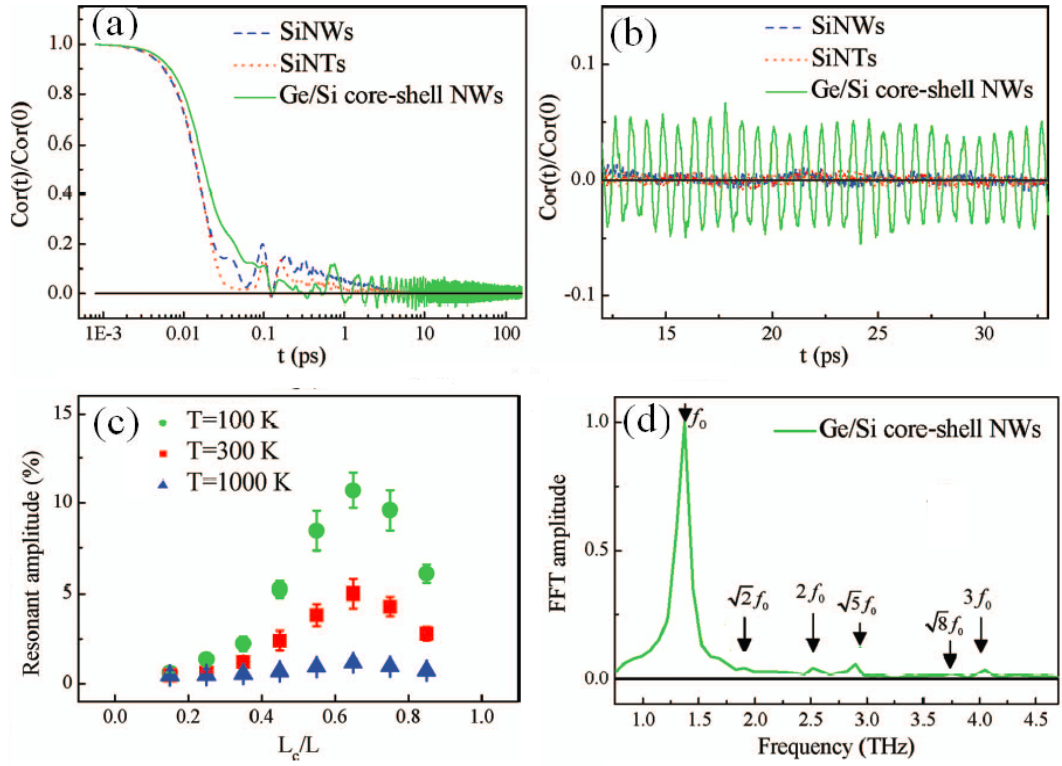

Fig.5 (Color online) (a) Time dependence of normalized heat current autocorrelation function (HCACF) for SiNWs (dashed line), SiNTs (dottedline), and Ge/Si core-shell NWs with Lc/L = 0.65 (solid line). (b) Long-time region of (a). (c) Oscillation amplitude versus core-shell ratio Lc/L. (d) Amplitude of the fast Fourier transform of the long-time region of normalized HCACF of Ge/Si core-shell NWs.

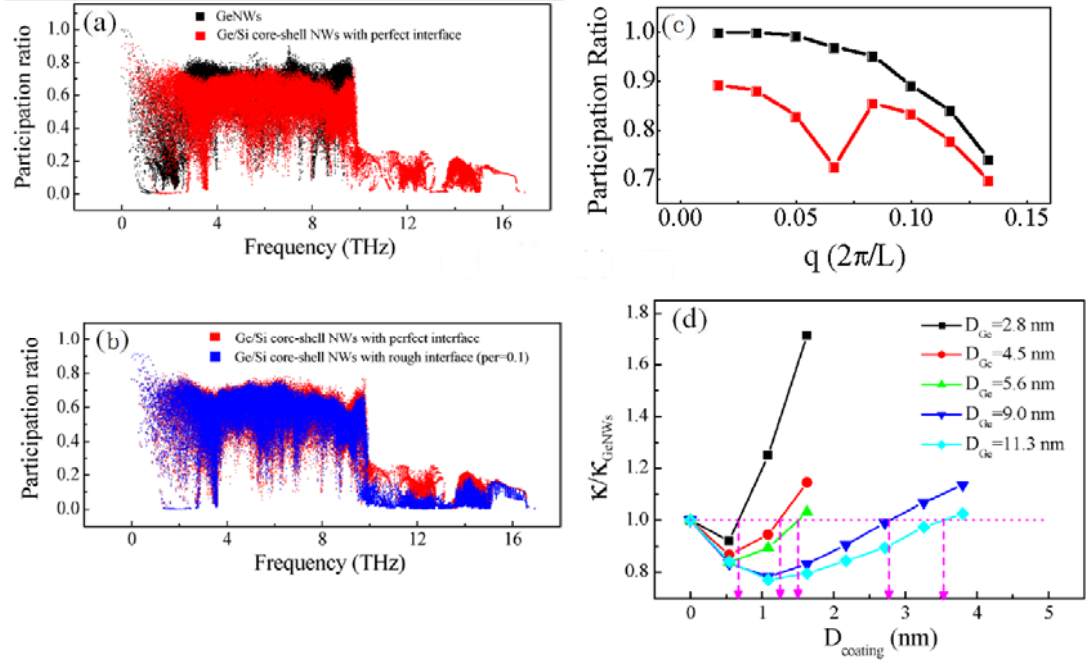

Fig.6 (Color online) (a) and (b) P-ratio for different phonon modes in GeNWs before and after coating. The black, red, and blue denote respectively p-ratio in GeNWs, Ge/Si core−shell NWs with perfect interface, and Ge/Si core−shell NWs with 10% interfacial roughness. (c) The polarization-resolved p-ratio for the longitudinal acoustic phonon near the Brillouin zone center in GeNWs (black) and Ge/Si core−shell NWs with perfect interface (red). (d) Normalized thermal conductivity versus coating thickness for different $D_{Ge}$. Thermal conductivity of GeNWs at each $D_{Ge}$ is used as reference.

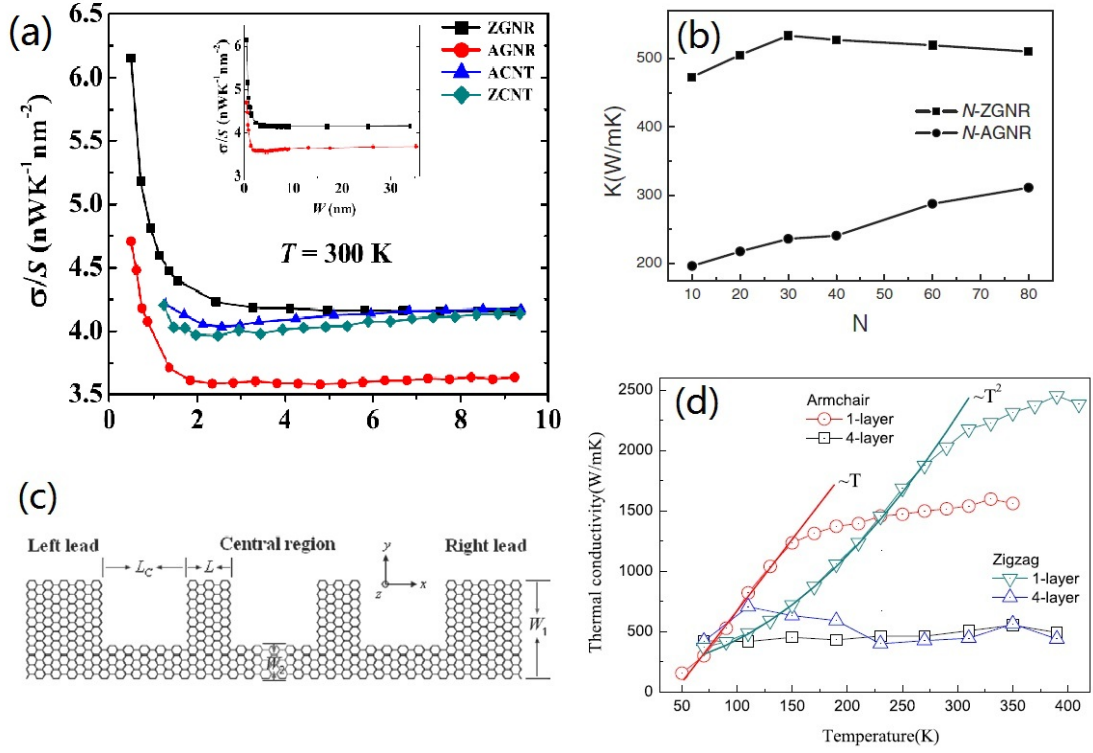

Fig.7 (Color online) (a) The scaled thermal conductance, σ/S at 300 K vs width, W, for Zigzag graphene nanoribbon (ZGNR), armchair GNR (AGNR), zigzag carbon nanotube (ZCNT) and armchair CNT (ACNT). The inset shows σ/S for ZGNRs and AGNRs with the width varying from 0.5 to 35 nm. (b) Thermal conductivity of N-AGNR and N-ZGNR with variation of N, where the length of GNRs is fixed to be 11 nm. (c) Schematic illustration of the periodic T-shaped GNR. The left or right lead has perfect periodicity with uniform width $W_1$ along the ribbon axis and the central region consists of constrictions with size $L_C \times W_2$ and stubs with size $L \times W_1$. (d) Temperature dependence of thermal conductivity of ZGNRs and AGNRs for one and four layers atomic planes.

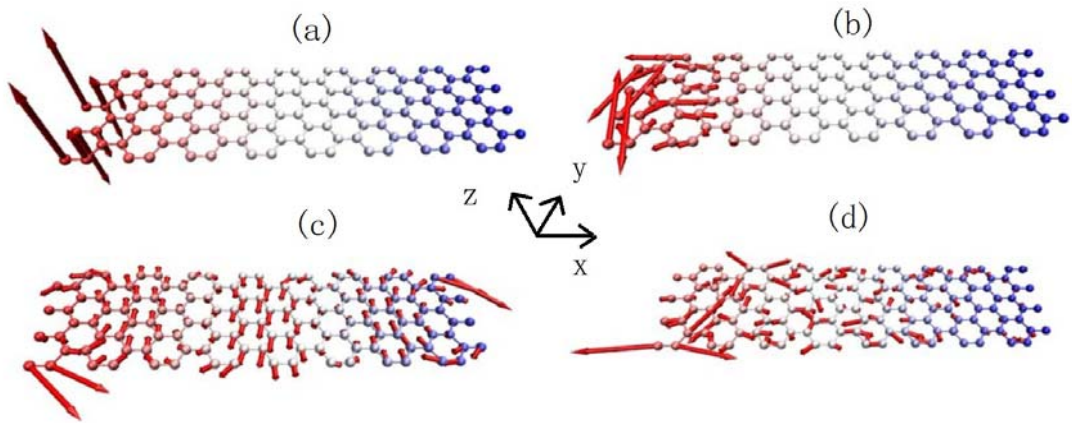

Fig.8 (Color online) (a) and (b) are two localized edgemodes in the graphene sheet. (c) and (d) are two non-localized modes due to the broken of different boundary conditions in three directions.

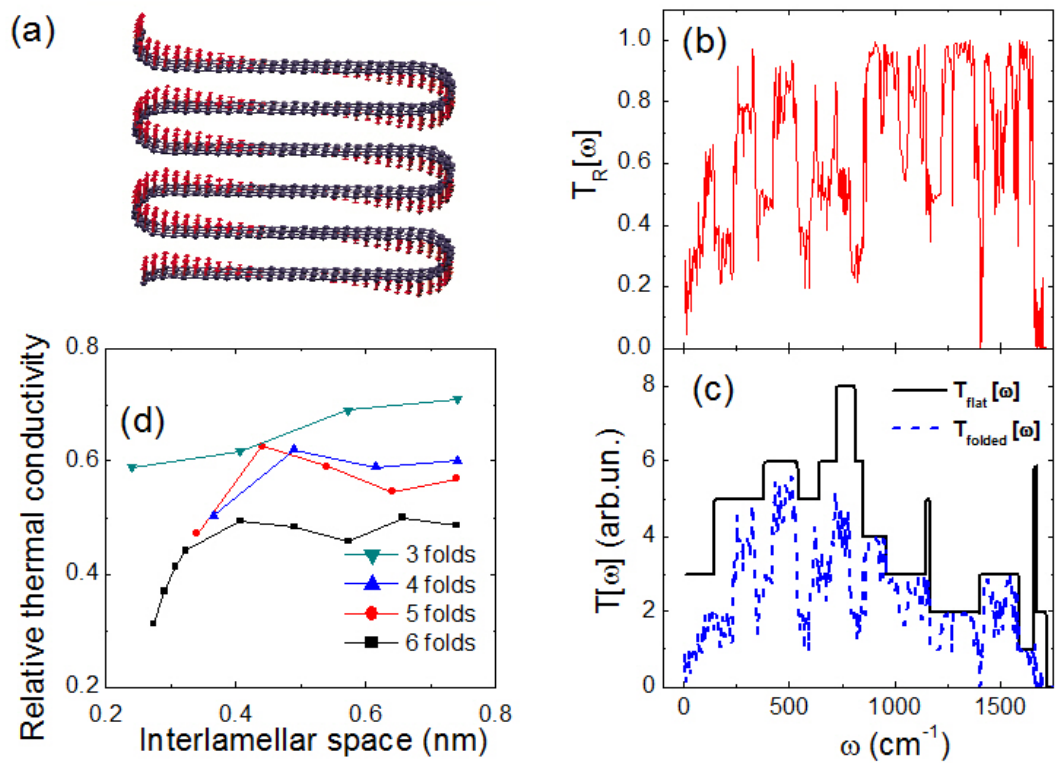

Fig. 9 (color online) (a) The side view of schematic picture of a folded GNR. Arrows correspond to eigenvectors of a ZA mode ($\omega$=50 cm$^{-1}$) in the folded GNR with periodic boundary condition in z-axis. The ZA mode is a combination of out-of-plane mode and in-plane mode. (b) The transmission ratio at a given frequency is the transmission coefficient of the folded GNR over the transmission coefficient of the flat GNR. It shows the scattering effect on the GNR by the folds. (c) The spectra of phonon transmission coefficient of the flat GNR and the folded GNR calculated by NEGF. (d) Relative thermal conductivity modulation by compressing interlamellar space with different folds in GNRs. The value 1.0 of relative thermal conductivity corresponds to 111.5 W/m-K which is the thermal conductivity of the flat zigzag GNR.

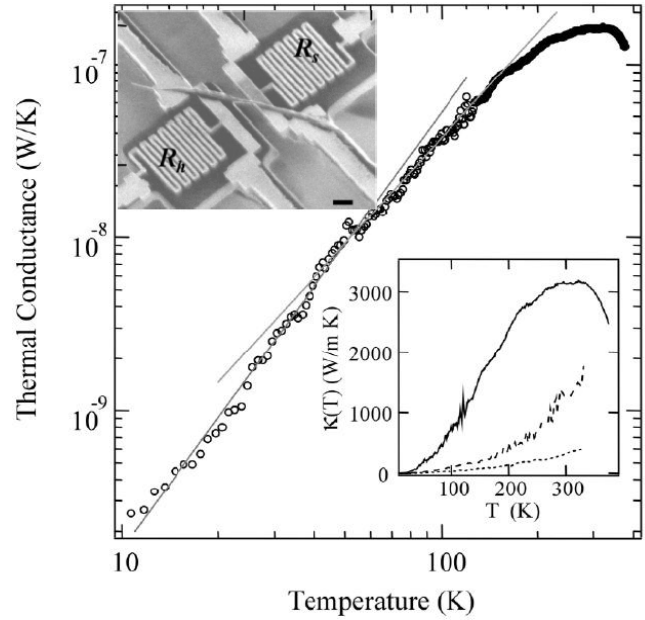

Fig. 10 Measured thermal conductance vs. temperature in an individual MWCNT with a diameter of 14nm. The slopes of the fitted solid lines are 2.50 and 2.01, respectively. Upper insert: SEM image of the two suspended membranes, $R_h$ and $R_s$, thermally connected by a MWCNT. The scale bar represents 1μm. Lower insert: thermal conductivity of three individual MWCNTs with different diameters (14nm, 80nm and 200nm from top).

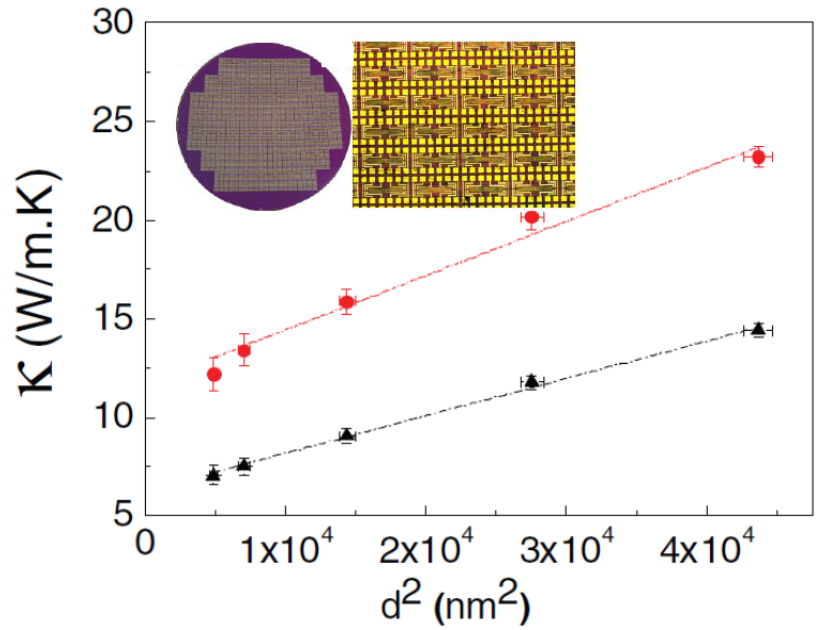

Fig. 11 (color online) Diameter dependence of thermal conductivity in ZnO nanowire at $T$ = 80K (solid circle) and 300K (solid triangles). The thermal conductivity increases linearly with cross-section area (~$d^2$) of nanowires. Insert: An 8-inch Si/SiN$_x$ wafer containing more than 600 MEMS test devices and zoom in image of the optical picture of 4×5 devices.

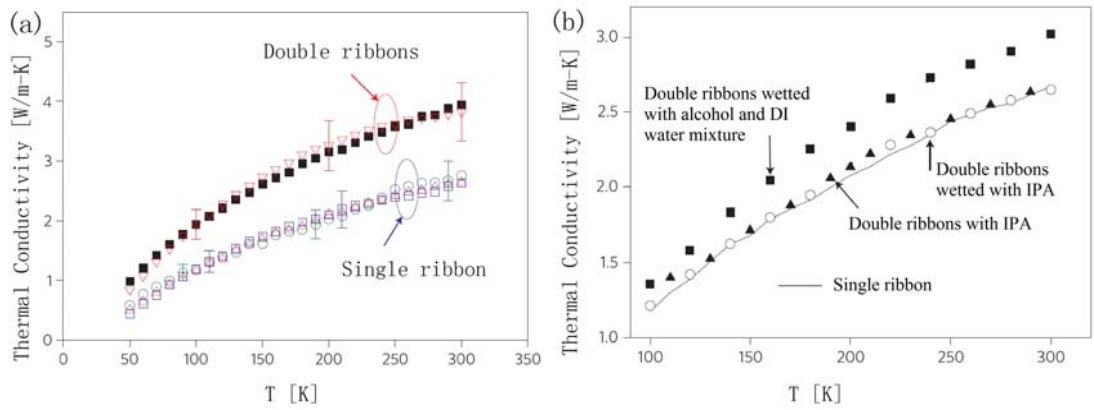

Fig. 12 (color online) (a) Measured thermal conductivity in double ribbons (coupled by van der Waals interaction) and single ribbon. The value in double ribbons is around 40% to 60% larger than that in single ribbon. (b) Switchable thermal conductivity of sample double ribbons with IPA, and reagent alcohol and DI water mixture wetting.

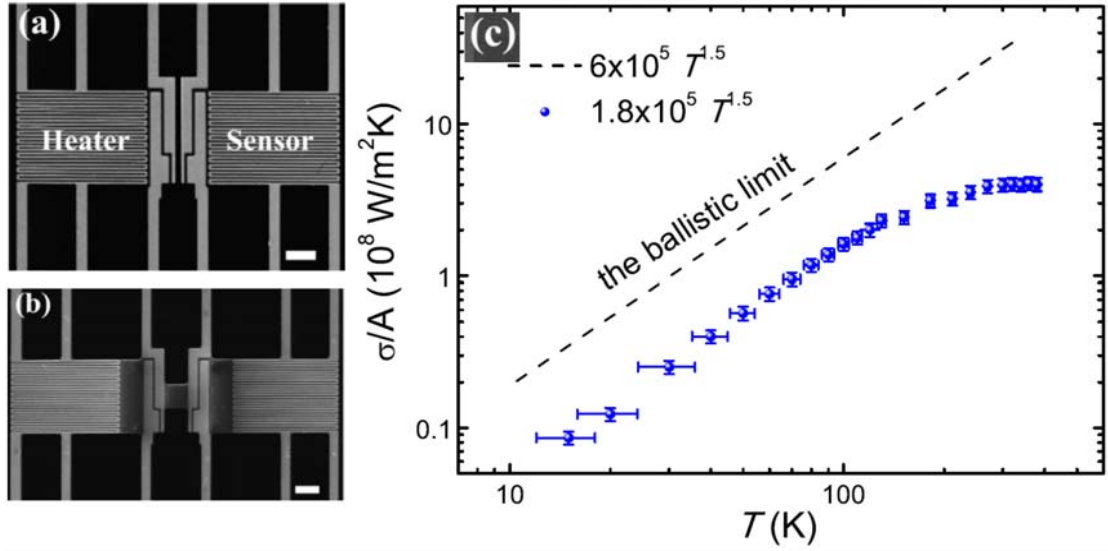

Fig. 13 (color online) (a) and (b) SEM image of suspended and supported graphene samples. Scale bar represents 5 μm. (c) Thermal conductance per unit cross section area σ/A in suspended single layer graphene. The measured data is approaching the expected ballistic limit (black dashed line).

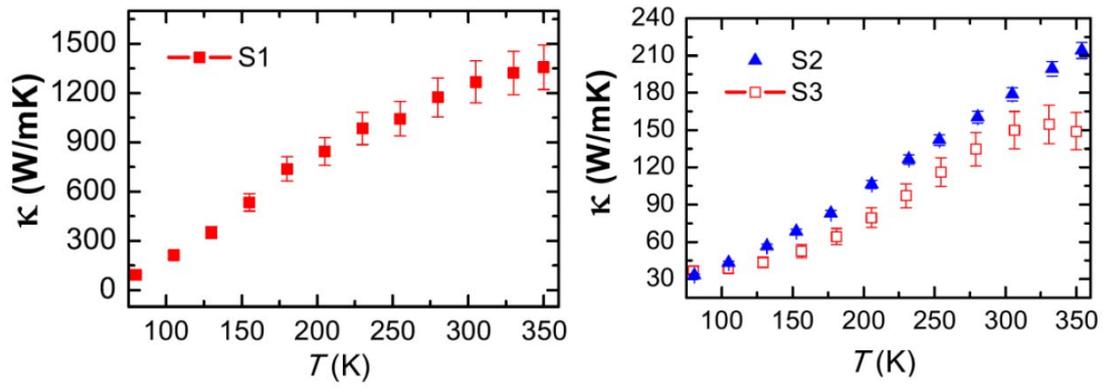

Fig. 14 (color online) Thermal conductivity of multilayer graphene as a function of temperature. The length of the samples S1 (supported three layers), S2 (suspended five layers) and S3 (supported three layers) are 5 μm, 2 μm and 1 μm, respectively, and the width is 5 μm.

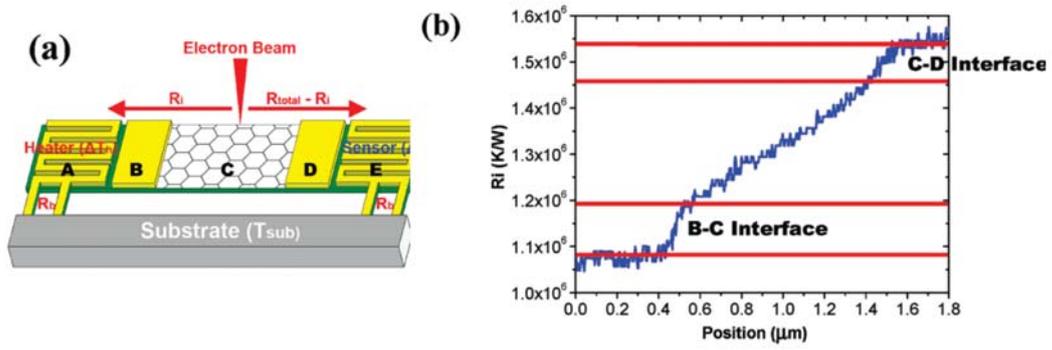

Fig. 15 (color online) (a) Schematic of the local electron heating technique to measure the thermal contact resistance. (b) Spatially resolved thermal resistance of supported graphene with 1μm in length. The sudden jump indicates thermal contact resistance in B-C and C-D interfaces.

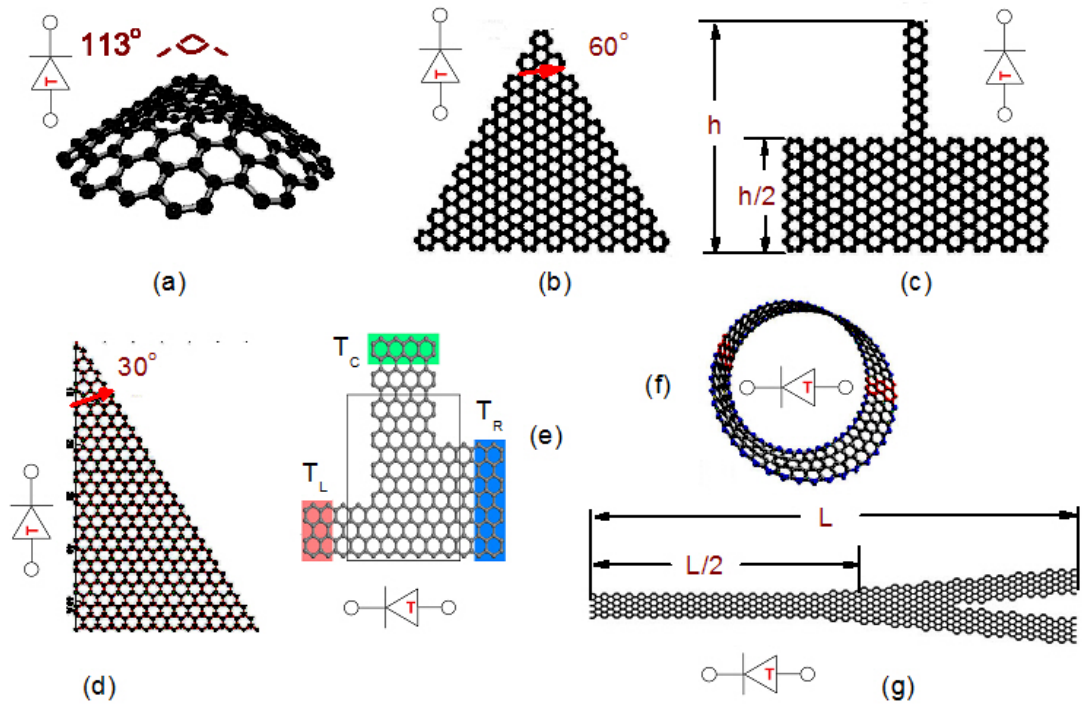

Fig. 16 (color online) Schematic pictures of thermal rectifier from different nanostructures. (a) The carbon nanocone. (b) The trapezia shaped GNR. (c) The two rectangular GNR with different widths. (d) The triangularly GNRs. (e) The asymmetric three-terminal GNR. Among the three terminals, the left and right terminals are energy-input or energy-output leads while the top terminal is a control lead. (f) The Möbius graphene strip with Zigzag edge and chiral index −1, where blue atoms are on the only edge. The red parts are heat bath regions. (g) The graphene Y junction. It consists of the stem section and branch sections, which have the same width. The symbol of diode with a red T repents a thermal diode.

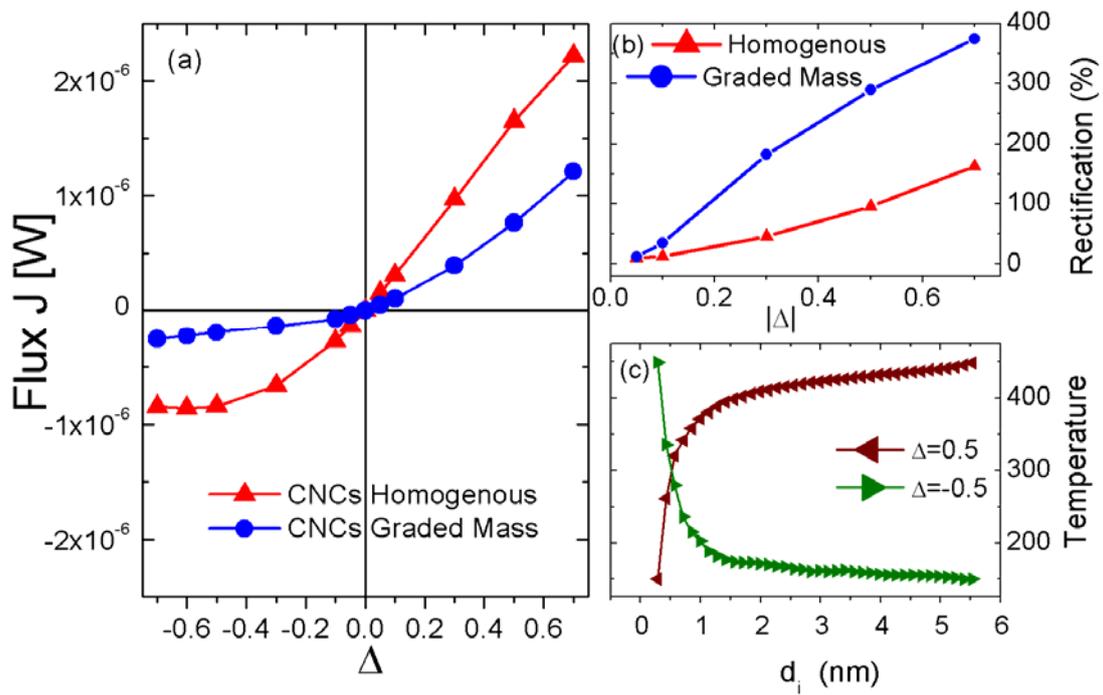

Fig. 17 (color online) (a) Heat flux J versus |Δ| for the homogenous mass and the graded mass carbon nanocones (CNCs). The temperature of top is $T_{top} = T_0(1-\Delta)$ and that of bottom as $T_{bottom} = T_0(1+\Delta)$, where $T_0$ is the average temperature, and Δ is the normalized temperature difference between the two ends. (b) Rectiffications versus Δ. (c) Temperature profile in CNCs at $T_0$=300 K and $\Delta = \pm 0.5$.

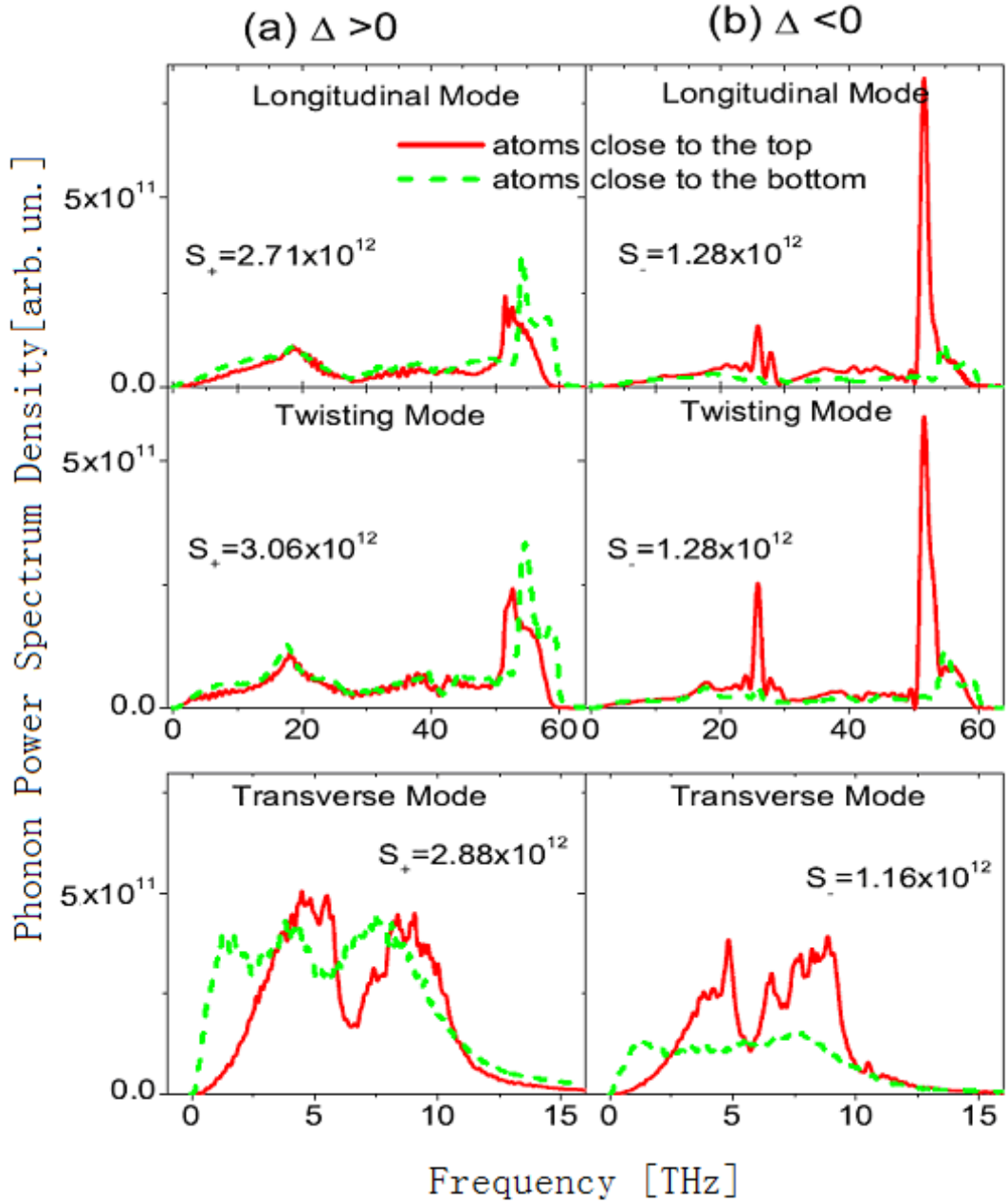

Fig. 18 (color online) Phonon power spectra of top/bottom atoms in the carbon nanocones. (a) $T_0$ = 300 K and $\Delta$ = 0.5, the bottom heat bath is at high temperature which corresponds to big flux; (b) $T_0$ = 300 K, $\Delta$ = -0.5, corresponds to small flux. The values of overlap area were shown in each panel. The values of $S_{+/-}$ (overlaps of the power spectra of the two layers) are also shown.

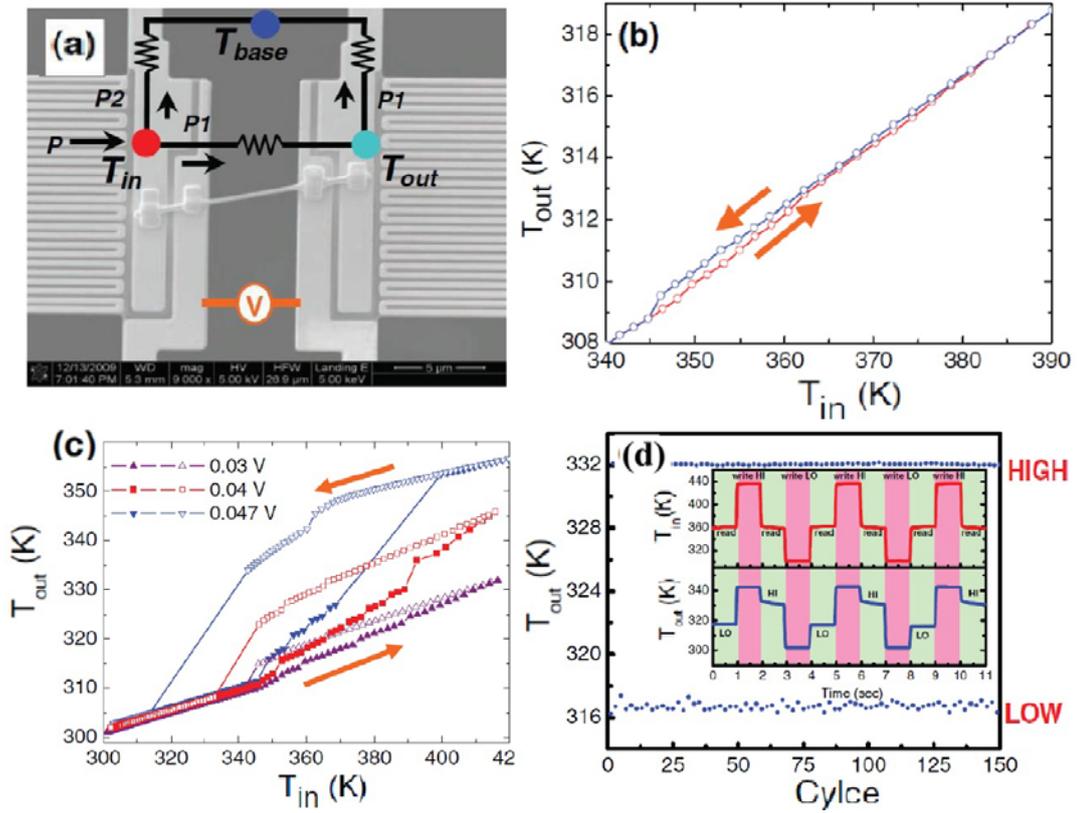

Fig. 19 (color online) (a) SEM image of a solid-state memory device with an individual $VO_2$ nanobeam as a tunable thermal channel connecting the input terminal ($T_{in}$) and output terminal ($T_{out}$). (b) $T_{out}$ as a function of $T_{in}$ within a single temperature-sweeping loop of heating (red curve) and cooling (blue curve). (c) $T_{out}$ as a function of $T_{in}$ under different bias voltage on $VO_2$ nanobeam. (d): Switching performance and repeatability test of the thermal memory within 150 cycles, each of which consists of a Write High-Read-Write Low-Read loop.